\documentclass[preprint2]{aastex}

\usepackage{bm,subfigure}

\shorttitle{Shear Estimates Using Neural Networks}
\shortauthors{Gruen, Seitz, Koppenhoefer, Riffeser}

\begin{document}


\title{Bias-Free Shear Estimation Using Artificial Neural Networks}


\author{D. Gruen\altaffilmark{1}, S. Seitz\altaffilmark{1,2}, J. Koppenhoefer\altaffilmark{1,2} and A. Riffeser\altaffilmark{1,2}}
\email{dgruen@usm.uni-muenchen.de}


\altaffiltext{1}{University Observatory Munich, Scheinerstrasse 1, 81679 Muenchen, Germany}
\altaffiltext{2}{Max Planck Institute for Extraterrestrial Physics, Giessenbachstrasse 1, 85748 Garching, Germany}


\begin{abstract}
Bias due to imperfect shear calibration is the biggest obstacle when constraints on cosmological parameters are to be extracted from large area weak lensing surveys such as Pan-STARRS-3$\pi$, DES or future satellite missions like Euclid.

We demonstrate that bias present in existing shear measurement pipelines (e.g. KSB) can be almost entirely removed by means of neural networks. In this way, bias correction can depend on the properties of the individual galaxy instead on being a single global value. We present a procedure to train neural networks for shear estimation and apply this to subsets of simulated GREAT08 RealNoise data.

We also show that circularization of the PSF before measuring the shear reduces the scatter related to the PSF anisotropy correction and thus leads to improved measurements, particularly on low and medium signal-to-noise data.

Our results are competitive with the best performers in the GREAT08 competition, especially for the medium and higher signal-to-noise sets. Expressed in terms of the quality parameter defined by GREAT08 we achieve a $Q\approx$ 40, 140 and 1300 without and 50, 200 and 1300 with circularization for low, medium and high signal-to-noise data sets, respectively.
\end{abstract}




\section{Introduction}

Weak gravitational lensing has proven to be a versatile method for measuring the mass distribution of galaxy clusters. With the detection of cosmic shear it has turned into an important tool for providing constraints on cosmological parameters such as $\sigma_8$ and $\Omega$ (see \citet{fu}).

Common to all applications of weak lensing studies is the requirement for statistical analysis of a great number of objects. Single sheared galaxies, due to their unknown intrinsic ellipticity and further observational uncertainties, can only give a reasonable shear estimate as part of a large sample. The accuracy of shear estimation methods poses a bottleneck to observational cosmology. Especially as surveys are getting larger (Pan-STARRS-3$\pi$, DES, Euclid), shear calibration bias could annihilate the gain from improved statistics for larger galaxy samples.

For this reason several competitions for the calibration of shear measurement pipelines and the development of improved methods using simulated data have been hosted.\footnote{cf. STEP1 \citep{step1}, STEP2 \citep{step2}, GREAT08 \citep{great08results}} The elimination of biases among the many methods presented there has, however, only been partially successful.

In this paper we make use of artificial neural networks in order to improve the existing shear measurement pipeline introduced by \citet{ksb} (KSB) and further developed by \citet{luppino} and \citet{hoekstra}. We apply these to the data simulated by GREAT08 \citep{great08results} and show that existing biases can be almost entirely removed.

\section{Motivation}

The fundamentals of the KSB method are to measure galaxy shapes derived from second moments $Q_{ij}$ integrated within a circular Gaussian weight function. From these, \emph{polarizations} can be defined as
\begin{equation}
\bm{e}:=\frac{1}{Q_{xx}+Q_{yy}} \left(\begin{array}{c} Q_{xx}-Q_{yy} \\ 2Q_{xy} \end{array}\right) \; .
\end{equation}

Observed polarizations $\bm{e}^{\mathrm{obs}}$ must be corrected for PSF anisotropy $\bm{p}$ and their responsitivity to shear $\bm{g}$ as based on PSF size and the galaxy's shape. KSB achieves this by linear corrections, such that
\begin{equation}
\bm{e}^{\mathrm{obs}}=\bm{e}^{\mathrm{true}}+P^{\mathrm{sm}} \bm{p} + P^{\gamma}\bm{g} \; ,
\label{eqn:ksb1}
\end{equation}
where $P^{\mathrm{sm}}$ is the galaxy's smear polarizability tensor and $P^{\gamma}$ is calculated as
\begin{equation}
P^{\gamma}=P^{\mathrm{sh}} - P^{\mathrm{sm}} \frac{P^{\mathrm{sh},\star}}{P^{\mathrm{sm},\star}} \; 
\label{eqn:ksb2}
\end{equation}
from smear polarizability tensors and shear polarizability tensors $P^{\mathrm{sh}}$ measured on the galaxy and the PSF (denoted with a star). These tensors are weighted fourth order moments of the respective light distributions.

Assuming that galaxies show no intrinsic alignment on the sky, one can thus obtain a shear estimate as
\begin{equation}
\langle \bm{g} \rangle = \langle P^{\gamma^{-1}} (\bm{e}^{\mathrm{obs}} - P^{\mathrm{sm}} \bm{p}) \rangle \; .
\label{eqn:ksb3}
\end{equation}

There exist a large number of implementations of KSB, differing from each other in subtle choices of the method for source extraction, determination of the radius for the weight function $r_g$, $P^{\gamma}$ tensor inversion, weighting, cuts (e.g., eliminating objects with large ellipticities or small values of $P^{\gamma}$), correction factors and further details \citep{step1}. It is not clear a priori and will likely depend on the particular data set which is the right choice on each of these points. Different decisions imply different biases (see STEP2, \citet{step2}), which have to be taken into account.

The bias introduced by not having calibrated a method correctly can only be avoided by carefully simulating the process for the very data that is to be analyzed.
This means that data with known shear have to be simulated. It must be the aim of these simulations to reproduce the properties of the respective sample of galaxies as accurately as possible (e.g., in terms of intrinsic light distribution, PSF, errors introduced by the data reduction and noise properties) in order for the calibrations to be appropriate.

In some cases using simulations a shear calibration bias has been found and corrected manually. For instance, T. Schrabback and \citet{mcinnes} multiply their shear estimates by $1/0.91$ and $1/0.82$, respectively, with factors found by calibration for STEP1 and STEP2 data. After such manual corrections, however, it is very likely that there is a residual bias, not only because the corrections done are usually very simple but also because bias likely differs for different galaxy properties and data sets.

We propose to make best use of the anyhow required simulations by letting neural networks analyze the data and estimate shear after training them on simulated data with known shear. Using a pipeline's ellipticity estimates and further parameters that might be indicative for the bias present on the respective galaxy, we test how well neural networks are able to eliminate biases and improve the shear estimate. Provided that such a scheme is successful, correct simulations allow for optimal calibration of shear measurement pipelines.

We point out that it is not the subject of this work to use neural networks on the pixelated light distribution of the galaxies itself, but to start from data which are quite close to an exact shear measurement already. The advantage of this method is that the network is fed with the most relevant parameters for shape estimates on a catalog basis, keeping training and application comparatively computation inexpensive. 

\section{Neural networks}

\begin{figure}[ht]
\centering
\includegraphics[width=0.48\textwidth]{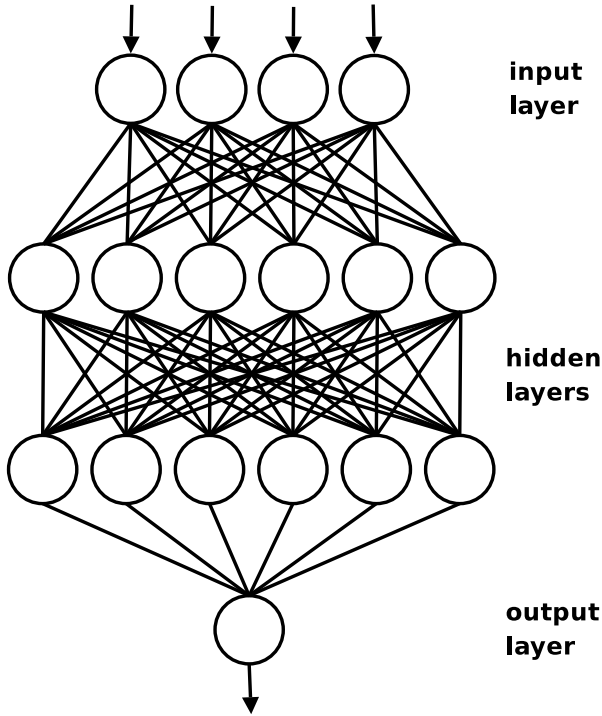}
\caption{sketch of a Perceptron with two hidden layers}
\label{fig:perceptron}
\end{figure}

Neural networks are nowadays commonly used in astronomy, be it for the detection and classification of objects or for finding photometric redshift estimates \citep{annz}. The flavor of networks most frequently used and also to be employed in this paper is multi-layer Perceptrons. From inputs $a^{\mathrm{in}}_i$ fed to an input layer of neurons, parameters are transferred through a number of hidden layers to finally make for one or more network outputs $x^{\mathrm{out}}_i$ (see Figure~\ref{fig:perceptron}). 

A neuron $i$'s output $x_i$ depends on the incoming signal from connected nodes $j$ in the previous layer, weighted by connection weights $w_{ji}$ and transformed by the neuron's activation function $f_i$:
\begin{equation}
x_i = f_i \left( \sum_j w_{ji} x_j \right) \; .
\end{equation}

While in general $f_i$ could be chosen differently for each node, it is usually taken to be the same nonlinear function for all hidden nodes and the identity for the input and output layers' nodes. In our case we use a sigmoidal function $f_i(a)=f(a)=(e^{-a}+1)^{-1}$, where node $i$ is a hidden node. The weights $w_{ij}$ of the connections between two adjacent layers' nodes $i$ and $j$ and from an additional bias node which accounts for an individual node threshold are to be optimized such that a cost function $E$ of the network output becomes minimal. A typical choice for the cost function is the sum over squared errors of outputs $x^{\mathrm{out}}_{k}$ on training sets $k$, for which true answers $\hat{x}_{k}$ are known. An additional term quadratic in the weights is added for regularization, penalizing large weights which characterize overfitting to specific data. Thus $E$ becomes
\begin{equation}
E=\sum_{k} (x^{\mathrm{out}}_{k} - \hat{x}_{k})^2 + \alpha \sum_{i,j} w_{ij}^2 \; .
\end{equation}

For training such a network, true answers for each training set have to be known, such that error back-propagation \citep{rum} can be used for optimizing the weights.\footnote{see appendix for a description of the algorithm, section~\ref{sec:bp}} 

However, for weak lensing measurements we can only expect the network to return a true shear component $g_l=:\hat{x}_l$ \emph{on average} for a large sample $l$ of galaxies. Training with true shear as the expected network output for all single galaxy data is counterproductive. We would rather like the networks to minimize the squared error between $g_l$ and $\langle e_{k}\rangle_k^l=:\langle x^{\mathrm{out}}_{k} \rangle_k^l$, i.e. the average output of galaxies $k$ on a sample of galaxies $l$, for each shear/ellipticity component.  We can express this with a cost function of the form
\begin{equation}
E=\sum_{l} (\langle x^{\mathrm{out}}_{k} \rangle_k^l - \hat{x}_{l})^2 + \alpha \sum_{i,j} w_{ij}^2 \; .
\label{eqn:costfunction}
\end{equation}
The back-propagation algorithm must in this case be adapted accordingly. We start from code provided by \citet{annz} and implement the algorithm as described in the appendix (section~\ref{sec:bpap}).

\section{Application to GREAT08 data}

For training and testing our network with the scheme described in more detail in the appendix, we use simulated galaxy images with known shear from the "RealNoise Blind" data sets of GREAT08 \citep{great08}.

\subsection{Sample selection}
\label{sec:samples}
Table~\ref{tbl:samples} gives an overview of the six samples analyzed in our work. Each of the samples contains sets of six image files of 10,000 simulated galaxies, each set being sheared with a true shear $(g_1,g_2)\in[-0.05,0.05]^2$ as plotted in Figure~A1 of \citet{great08results}.

From the 2700 sets of galaxies in the "RealNoise Blind" sample of GREAT08, we pick those 1500 sets from the fiducial (medium) signal to noise group which share the same PSF (the fiducial PSF also labeled PSF 1 by GREAT08) but differ in terms of galaxy size and type. In the following analysis we denote these as sample 1.

In order to find how well such corrections can work on galaxies with different signal-to-noise levels, we pick two more samples. As these are homogeneous in all galaxy property distributions, they are not as realistic as sample 1 but still can give an indication of the dependence of network performance on signal-to-noise ratio. Therefore, we perform a similar analysis with the 300 high signal to noise data sets (sample 2) and the 300 low signal to noise sets (sample 3), which all share the same galaxy properties and PSF. It should be noted, though, that in the latter case signal is so low that without an input catalog source extraction suffers a significant rate of false detections. On real single frame data with similar signal to noise ratio, the centroid position on stacked frames could be used to eliminate false detections and have well-defined centroid positions, improving the accuracy of the measurement. In our case we cross-correlate catalogs against the GREAT08 grid positions to achieve complete and clean detections.

\subsubsection{Circularized samples}

In order to study the influence of PSF circularization as an alternative method of PSF anisotropy correction, we repeat the analysis with three more samples.

A sample of all galaxies with medium signal-to-noise level is denoted as sample 1c. Unlike sample 1, it contains data with all three PSFs used in the GREAT08 challenge. For this sample we circularize all three PSFs to the same circular target PSF before running KSB. We are using the method described by \citet{al} in its implementation by \citet{arno}. The target PSF is similar to the initial ones but slightly larger and circular, with a Moffat profile of 3 pixels FWHM and $\beta=3.2$. After having built a model of the initial PSF using 100 stars each, we convolve the data with a kernel model consisting of a superposition of four Gaussians with $\sigma=1,3,9,0.1$ multiplied with polynomials in $x$ and $y$ of order $n=6,4,2,0$, respectively. We do a $\chi^2$ fit of the 50 model parameters to reach our target PSF by convolution with this kernel. 

Due to the larger size of sample 1c, we reserve a larger number of galaxies for blind testing the networks later on. We prepare two more similarly circularized samples 2c and 3c using the data from samples 2 and 3. These data sets are homogeneous in all galaxy properties and therefore not as realistic as sample 1c.

\begin{deluxetable}{llllllll}
\tabletypesize{\scriptsize}
\tablecaption{Properties of the six samples used for neural network analysis, see section~\ref{sec:samples}\label{tbl:samples}}
\tablewidth{0pt}
\tablehead{
\colhead{sample} & \colhead{galaxy sets\tablenotemark{a}} & \colhead{s/n\tablenotemark{b}} & \colhead{PSF\tablenotemark{c}} &  \colhead{galaxy properties\tablenotemark{d}} & \colhead{gradient sets\tablenotemark{e}} & \colhead{validation sets\tablenotemark{e}} & \colhead{blind sets\tablenotemark{f}}
}
\startdata
\tableline
1 & 1500 & 20 & 1 & mixed    & 1394 & 53 & 53\\
1c\tablenotemark{g}& 2100 & 20 & 1/2/3 & mixed& 867 & 33 & 1200\tablenotemark{h}\\
2 &  300 & 40 & 1 & fiducial &  260 & 10 & 30\\
2c\tablenotemark{g} &  300 & 40 & 1 & fiducial &  260 & 10 & 30\\
3 &  300 & 10 & 1 & fiducial &  260 & 10 & 30\\
3c\tablenotemark{g} &  300 & 10 & 1 & fiducial &  260 & 10 & 30\\
\enddata
\tablenotetext{a}{each galaxy set contains 10000 galaxies}
\tablenotetext{b}{as defined by \citet{great08results}, the s/n estimate from the KSB$_S$ pipeline is considerably lower}
\tablenotetext{c}{using the PSF name convention as in \citet{great08}}
\tablenotetext{d}{GREAT08 simulate galaxy sets with different galaxy sizes and galaxy sets featuring either concentric bulge and disc, off-center bulge and disc or only one of either bulge or disc models; the fiducial group corresponds to medium size and concentric bulge and disc; as real data will typically contain a diversity of galaxy properties, samples 1 and 1c come closer to realistic samples}
\tablenotetext{e}{the sets used for training are split up randomly for each of the 500 differently initialized and trained networks into sets used for finding the gradient and sets used for validating the solution during the training process}
\tablenotetext{f}{these are galaxy sets put aside and not used for training, validating or selecting the networks}
\tablenotetext{g}{c stands for circularization; we have circularized all three anisotropic PSFs to the same circular PSF in this sample}
\tablenotetext{h}{a large number of blind sets have been reserved for extensive blind testing on this sample}
\end{deluxetable}

\subsection{Running KSB}

On the 2700 sets of 10000 galaxies each from the GREAT08 "RealNoise Blind" challenge we run a KSB implementation KSB$_S$, based on the version assembled by T. Schrabback and denoted as TS in \citet{step2}. (See also \citet{erben}).

After source extraction with SExtractor the pipeline uses \texttt{analyse} to calculate for each galaxy the tensors $P^{\mathrm{sm}}$ and $P^{\mathrm{sh}}$.\footnote{cf. eqn.~\ref{eqn:ksb1} to \ref{eqn:ksb3}} PSF anisotropy correction is done with this and $P^{\gamma}$ is computed as 
\begin{equation}
P^{\gamma}_{ij} = P^{\mathrm{sh}}_{ij}-\frac{\mathrm{tr}(P^{\mathrm{sh,\star}})}{\mathrm{tr}(P^{\mathrm{sm,\star}})}P^{\mathrm{sm}}_{ij} \; ,
\end{equation}
stars denoting quantities measured on the PSF.

This tensor is applied to the measured polarizations using trace inversion to find individual galaxy shear estimates 
\begin{equation}
\bm{e}^{\mathrm{iso}}=\frac{2(\bm{e}^{\mathrm{obs}} - P^{\mathrm{sm}} \bm{p})}{\mathrm{tr}(P^{\gamma})}
\end{equation}
Objects with $\mathrm{tr}(P^{\gamma})<0.1$ are discarded. The estimate labeled as KSB$_S$ in the following analysis is always $\bm{e}^{\mathrm{iso}}/0.91$, scaled with a calibration factor as optimized for this implementation of KSB using STEP1 data.

\subsection{Neural network training}

From the output of the KSB$_S$ pipeline we take $\bm{e}^{\mathrm{iso}}$ as the starting point for neural network analysis. As potential predictors for bias we add the weight function radius $r_g$ which in our pipeline is equal to SExtractor's \texttt{FLUX\_RADIUS}, the flux as measured by \texttt{analyse}, all four components of $P^{\gamma}$ and the pipeline's error estimates for the initial shape measurements $\Delta\bm{e}$.

The networks used feature three hidden layers of ten nodes each for both components and are trained using the algorithm described\footnote{see appendix, section~\ref{sec:bpap}} and the true shears published by GREAT08 after the end of the challenge. We split each sample into a subset used for training and a subset for later blind testing of network performance. Following the optimized ratio of gradient to validation sets derived by \citet{amari} for the asymptotic case of many available sets, we split the training sets again into subsets used for determining the gradient and others required for validation during training. The respective sizes of subsets are given in Table~\ref{tbl:samples}. For each sample the training process is iterated 500 times with different random initial weight configurations and a random allocation of training sets into gradient and validation sets.

\subsection{Selecting and blind testing networks}
It is necessary to ensure for evaluating the networks or any real-world application that the networks trained really perform consistently well on the data used for training and similar data not used for training or selecting the networks, in our case the blind sets (cf. Table~\ref{tbl:samples}).

Overtrained networks are generally characterized by some weights becoming comparatively large. While a penalization of large weights by the second term in eqn.~\ref{eqn:costfunction} already reduces overfitting to the training sample, where the number of training sets is sufficiently small overtraining may still occur because the reduction in errors from overfitting outweights the penalization due to large weights. For all following analyses, we therefore use the sum of squared weights,
\begin{equation}
S = \sum_{i,j} w_{ij}^2 \; ,
\end{equation}
to discard networks which are more than $1\sigma$ above the average in $S$ for the sample of networks cropped at 1.2 times the median $S$. This deselects about half of the networks on each of the samples, some of which might in fact not be overtrained. In the presence of larger samples, therefore, when more careful selection of networks is possible, performance might still increase. For all following analyses, we only use the weight-selected networks not discarded by these criteria.

\begin{figure}[htp]
\centering
\plotone{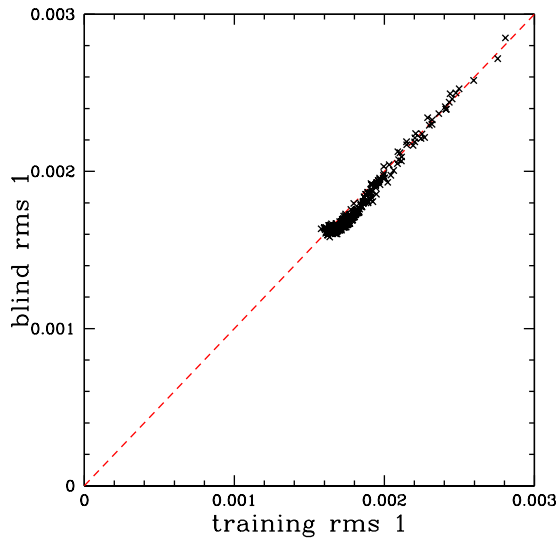}
\caption{Comparison of the performance (in terms of rms) on training data and blind data of the weight-selected networks trained on sample 1c, component 1. The dashed line is the identity.}
\label{fig:blindr1}
\end{figure}

To compare the performance of the networks left on training and blind data, we plot the root mean square error of the shear $g_i^o$ measured against the true shear $g_i^t$,
\begin{equation}
\mathrm{rms}=\sqrt{\langle (g_i^o - g_i^t)^2 \rangle} \; ,
\label{eqn:rms}
\end{equation}
which the weight-selected networks achieve on the data used for training and the blind sets. The plot shown in Figure~\ref{fig:blindr1} shows the result for sample 1c, component 1, for which the blind rms and training rms are equal within statistical uncertainty. For the other samples and components, due to the smaller number of blind sets, scatter is significantly larger and small constant offsets from the identity in both directions occur, likely due to the particular properties of the blind sets. In all cases, the networks performing best on the training data perform consistently well on blind data.

We select the best networks on each sample and component simply by taking the network with the smallest squared errors on the training data, not taking into account their performance on the blind data. After having discarded overfitted networks according to their weights as described above, this results in networks performing consistently well on training data and blind data. We perform the following analyses on the blind data sets only. As the results found on training and blind data agree within the statistical uncertainty, we use the complete sample of data sets for the analysis done in Section~\ref{sec:sixpack}, as this is necessary here. For sample 1c, also the analysis in Section~\ref{sec:sixpack} is done exclusively on the blind sample.

\subsection{Shear measurement performance}

We analyze the performance of plain KSB$_S$ and the neural networks selected in the previous section. For each sample and component, we calculate both for the blind and the training sets a quality parameter
\begin{equation}
Q_{1}=\frac{10^{-4}}{\mathrm{rms}^2}\;,
\label{eqn:q}
\end{equation}
averaging the rms (cf. eqn. \ref{eqn:rms}) over all galaxy sets within each sample. Resulting values of $Q_1$ are shown in Tables~\ref{tbl:bias} and \ref{tbl:biasc}. Note that $Q_1$ is smaller than the GREAT08 quality parameter $Q=:Q_6$ because we do not average the residuals over similar sets\footnote{That is, sets that have the same true shear, PSF, signal-to-noise and galaxy properties.} until section~\ref{sec:sixpack}.

Results of the circularized samples 1c to 3c are generally better in comparison to similar sets with anisotropic PSFs which have to be corrected for by a $P^{\mathrm{sm}} \bm{p}$ term.\footnote{cf. eqn.~\ref{eqn:ksb3}} A more detailed discussion of the advantages of circularization combined with bias correction is given in section~\ref{sec:circ}.

\subsection{Bias analysis I: linear bias}
\label{sec:linbias}

We calculate additive and multiplicative biases, following \citet{step1} and \citet{step2}. We apply a linear fit of residual shears $g_i^o-g_i^t$ against true shears $g_i^t$, i.e.
\begin{equation}
g_i^o-g_i^t=m_i\cdot g_i^t + c_i \; .
\end{equation}

\begin{figure*}
\centering
\epsscale{.50}
\subfigure[sample 1/1c (medium signal to noise)]{
\plotone{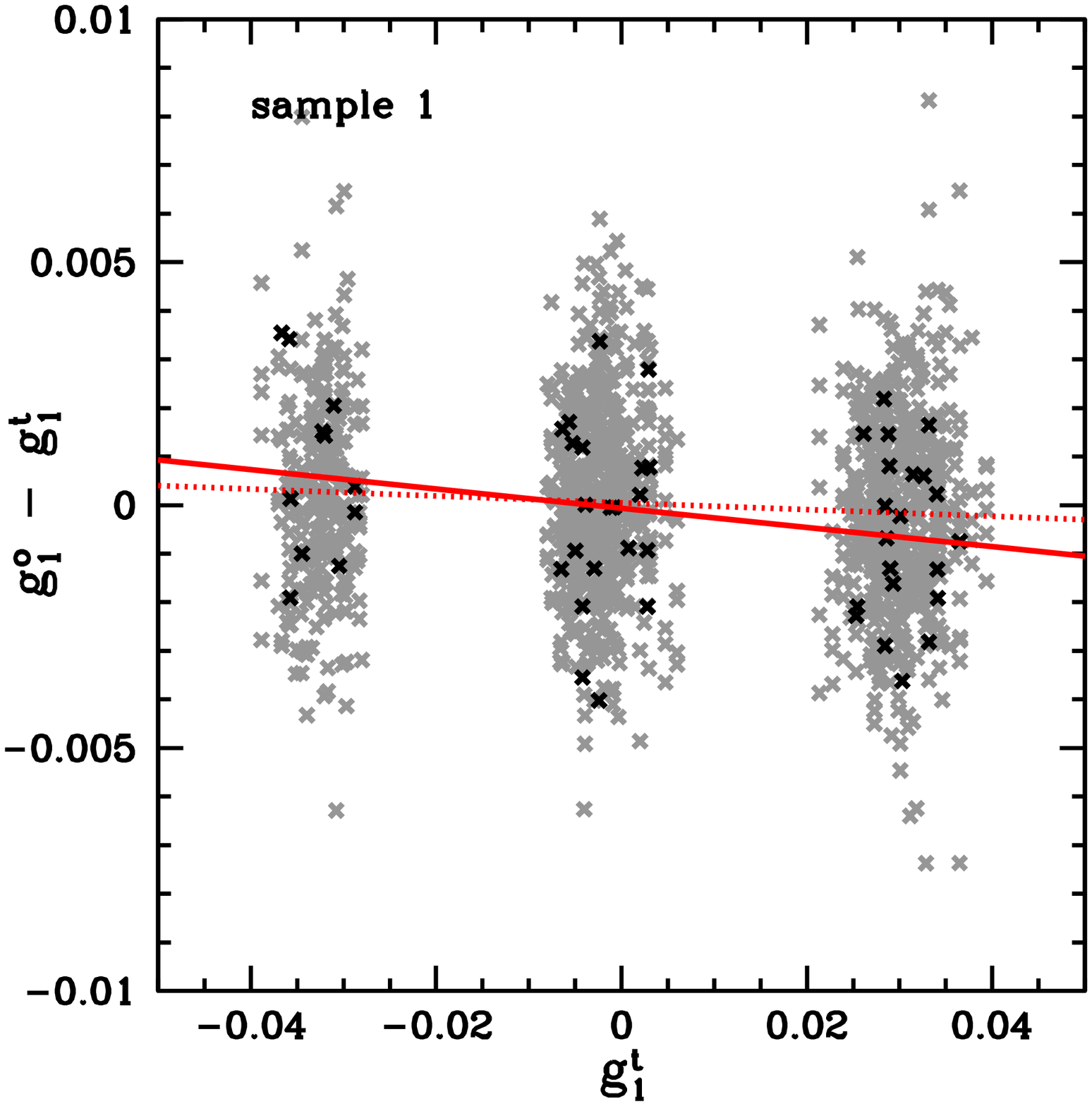}
\plotone{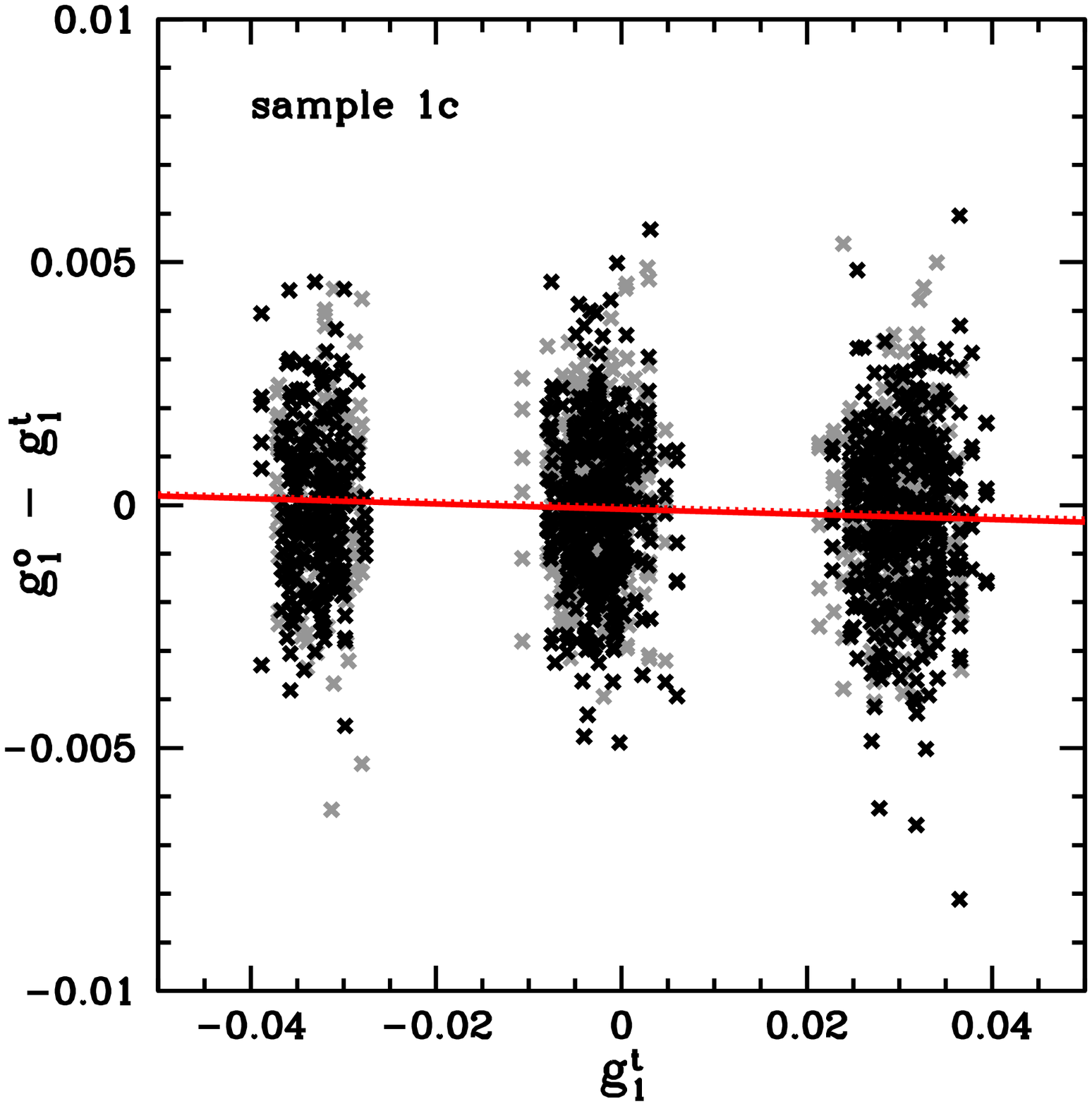}
}
\subfigure[sample 2/2c (high signal to noise)]{
\plotone{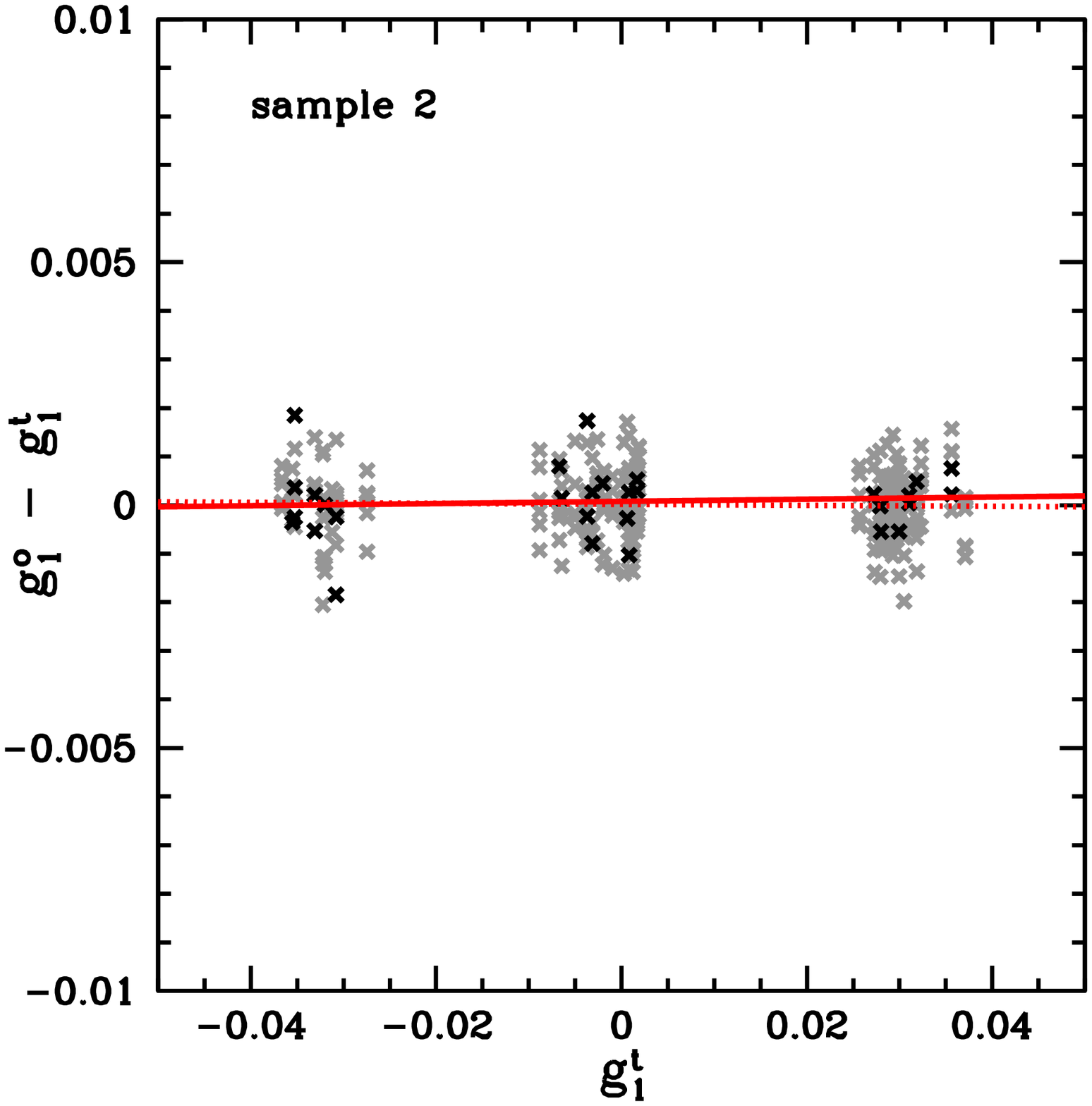}
\plotone{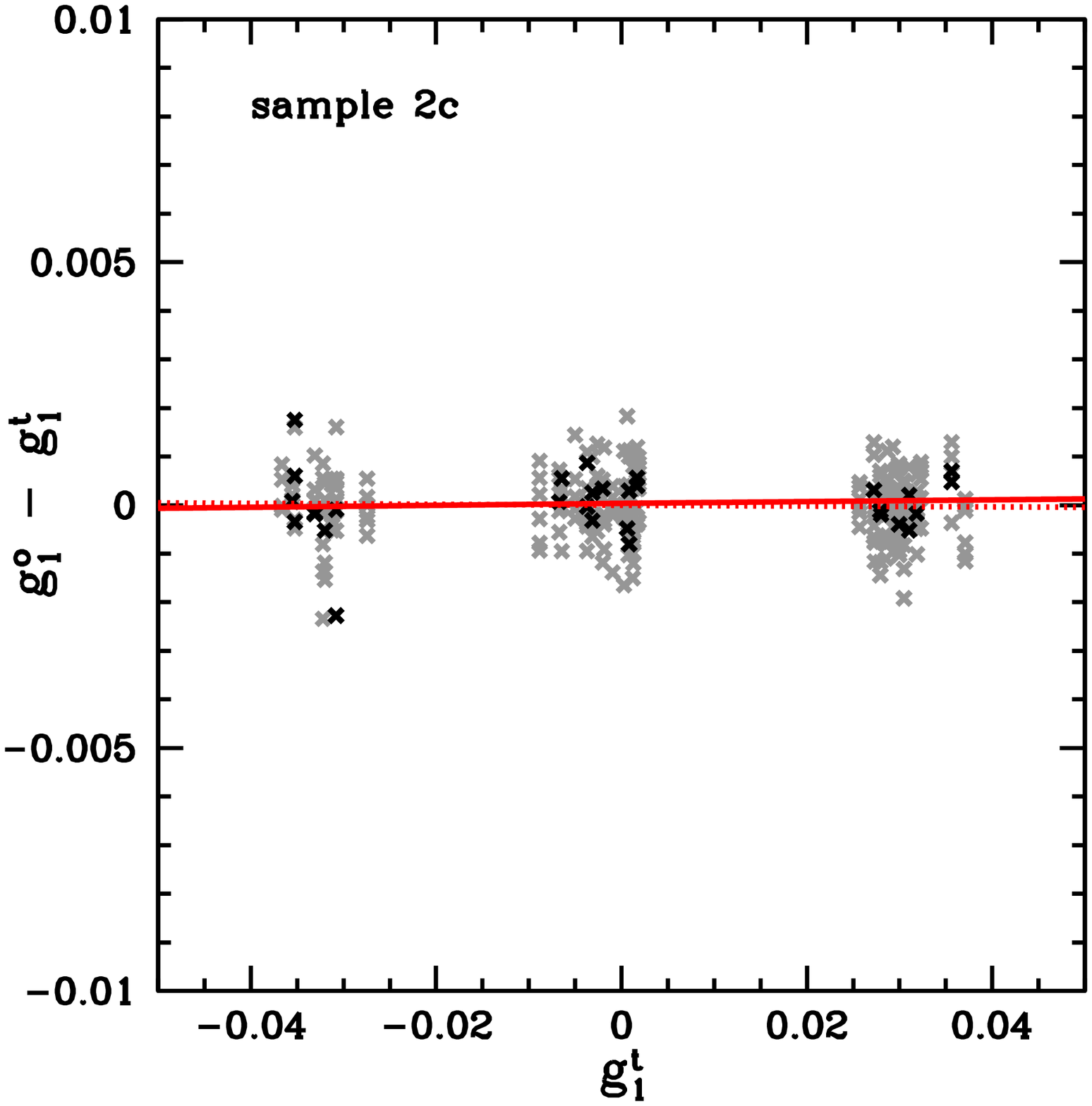}
}
\subfigure[sample 3/3c (low signal to noise)]{
\plotone{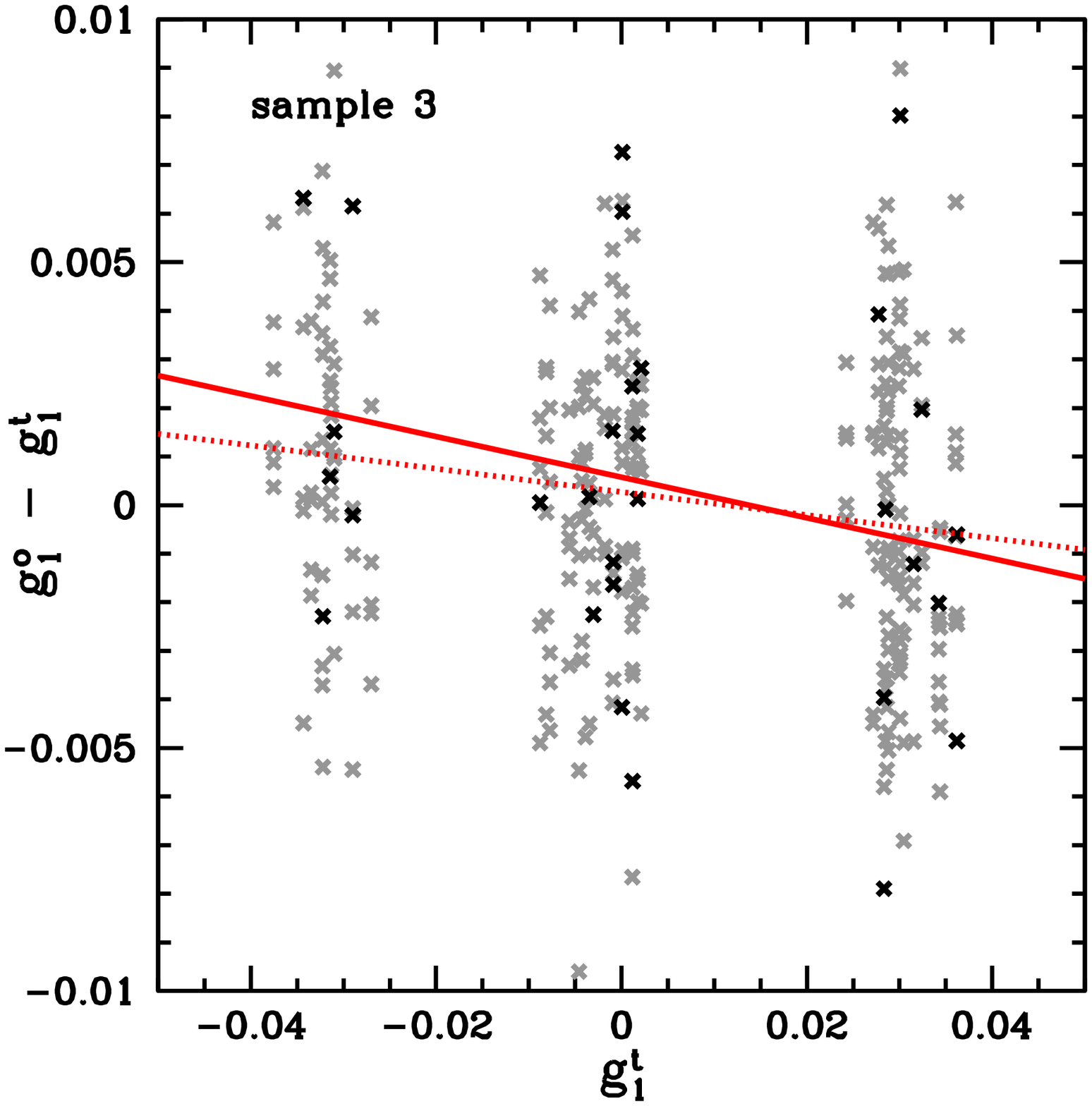}
\plotone{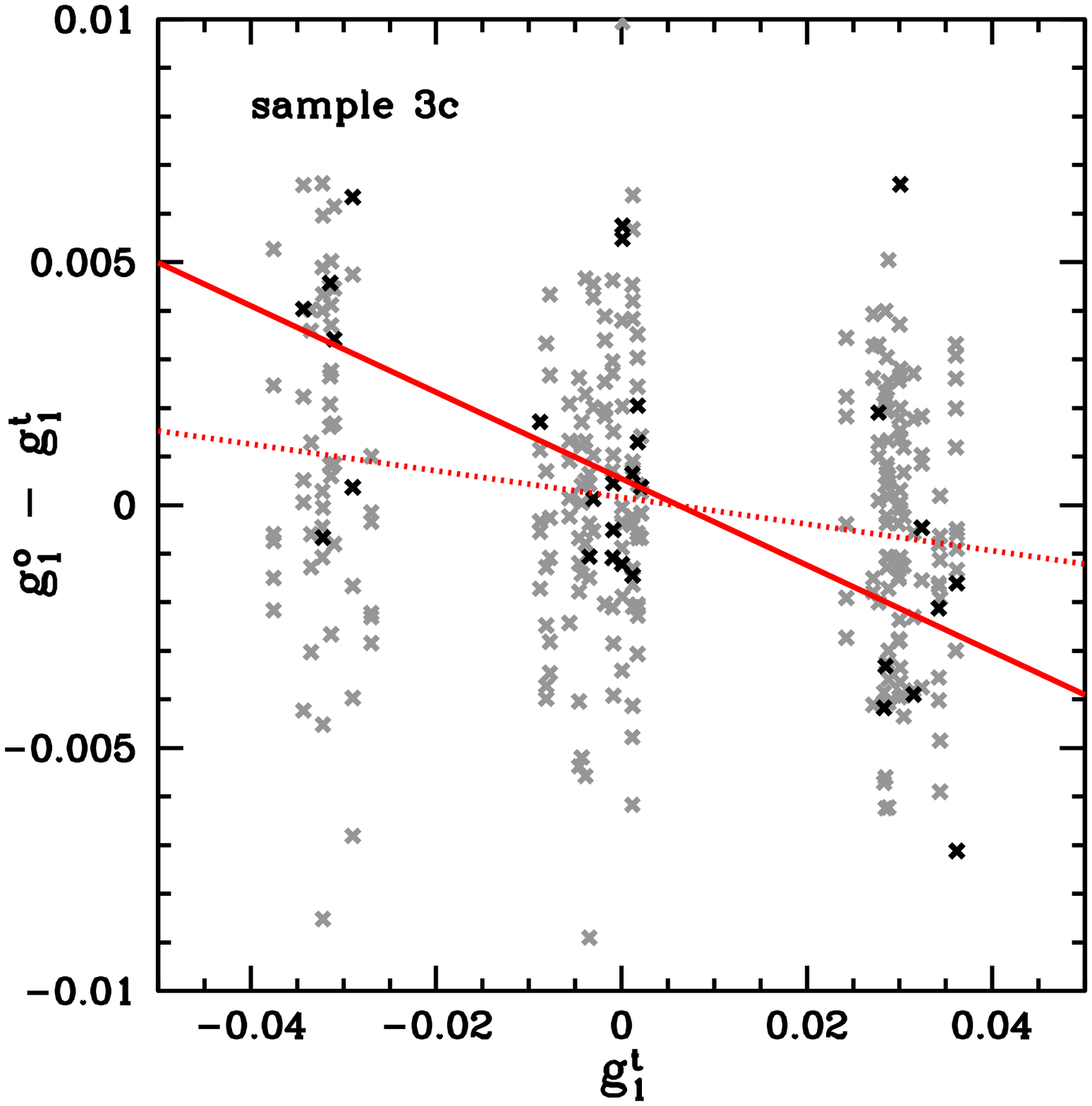}
}
\caption{Residual shear versus true shear for the neural network estimate KSB$_S$+NN on different data sets with linear fits for the additive and multiplicative shear measurement bias. Left panels: non-circularized data; right panels: circularized data. Plotted is always component 1, component 2 of the shear is very similar. Each data point corresponds to the shear estimate for one set of 10000 galaxies. Black points and solid lines correspond to blind data, grey points and dashed lines to all data.}
\label{fig:linbias}
\end{figure*}

Results for the six samples are shown in Tables~\ref{tbl:bias} and \ref{tbl:biasc} and Figure~\ref{fig:linbias}. For neural network analysis, both multiplicative and additive bias are well within the range of most successful methods in the GREAT08 competition \citep{great08results}. The requirements for future surveys as computed by \citet{amara} are always fulfilled for medium and high signal to noise in terms of $c_i^2<10^{-7}$. For the multiplicative bias criterion, $m_i<10^{-3}$, the network estimate is successful at least within an order of magnitude. The higher the signal-to-noise ratio, the better multiplicative bias can also be corrected, while especially for smaller signal there seems to be a tendency of $m<0$ for the neural network estimate, potentially due to the weaker shear signal.

\begin{figure*}
\centering
\subfigure[multiplicative bias]{
\includegraphics[clip, width=0.8\textwidth]{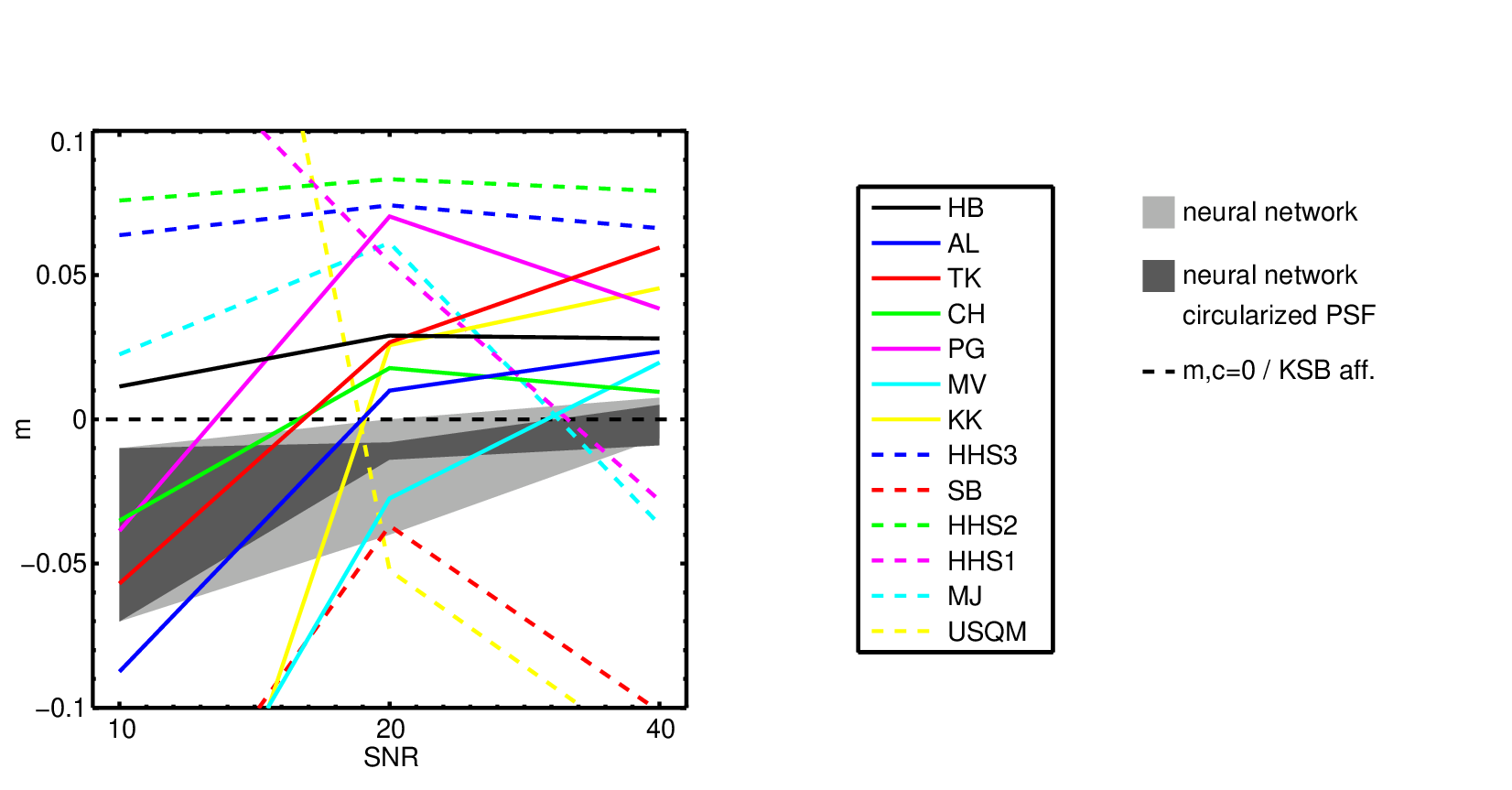}
}
\subfigure[additive bias]{
\includegraphics[clip, width=0.4\textwidth]{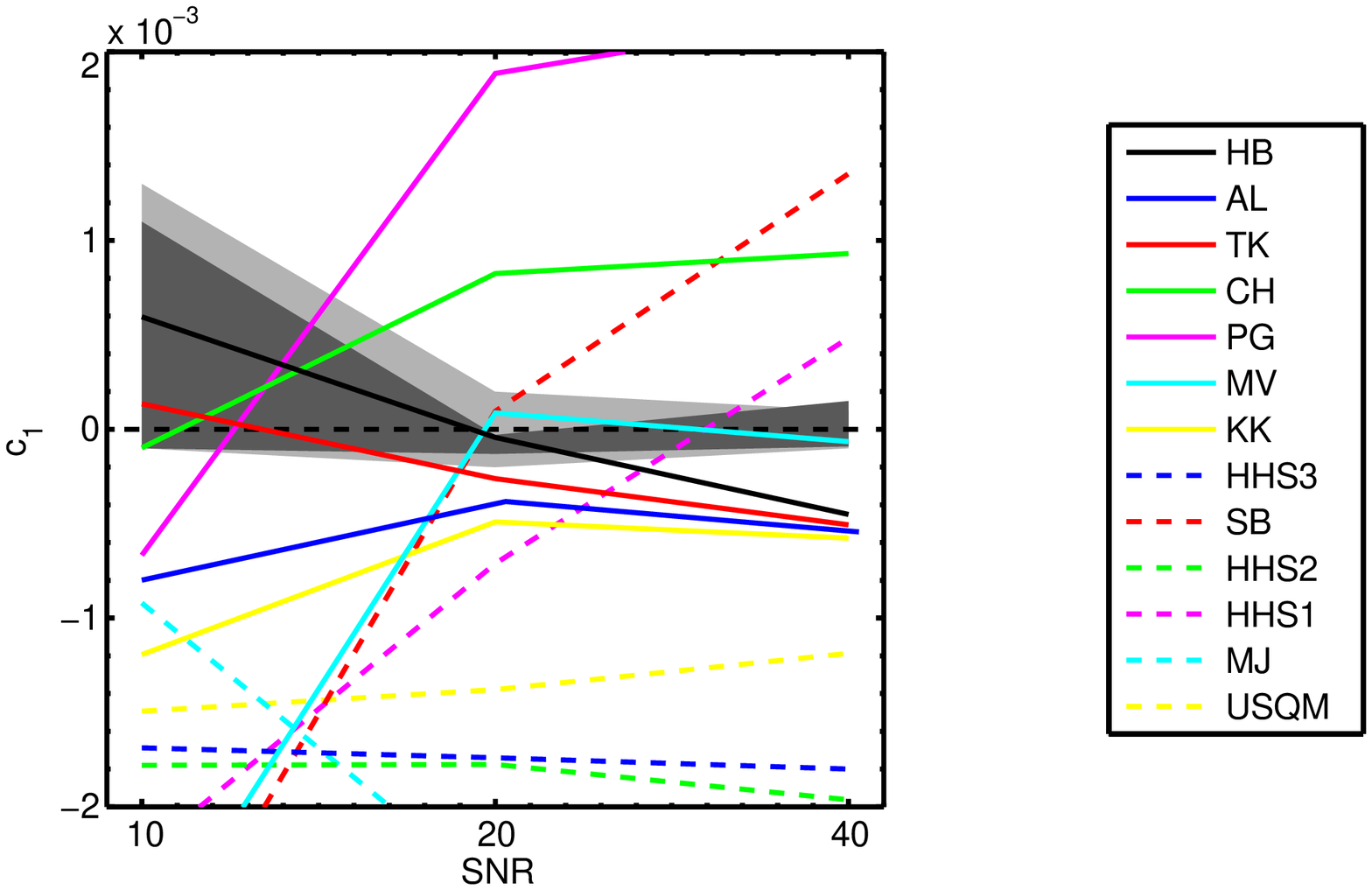}
\includegraphics[clip, width=0.4\textwidth]{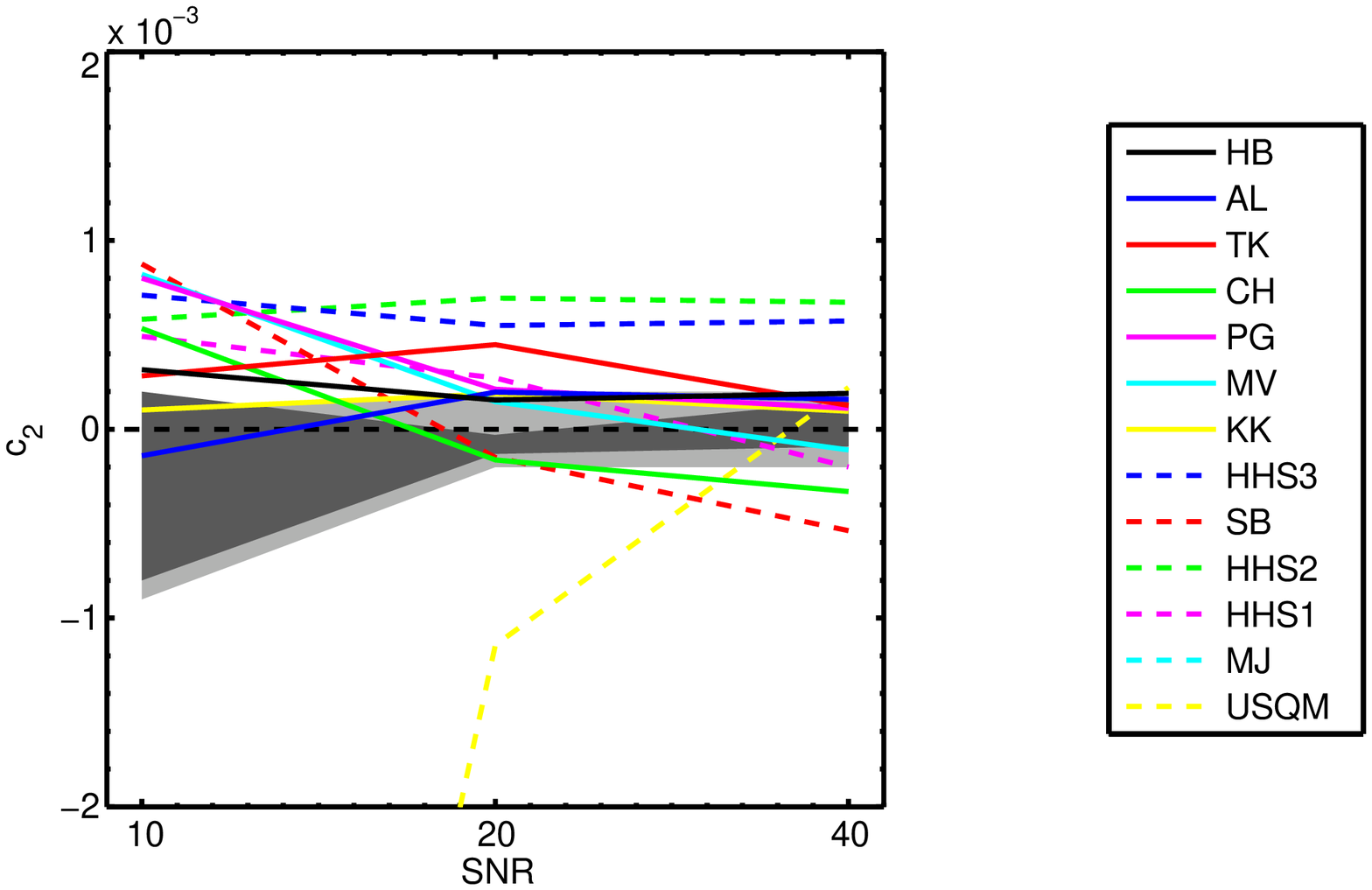}
}
\caption{multiplicative and additive bias as a function of signal-to-noise ratio for methods participating in the GREAT08 competition (see legend and cf. \citet{great08results}) and neural network estimates on the blind samples with (dark grey) and without (light grey) circularization; the dashed black line is at $m=c_i=0$, which is also the result for KSB aff. as defined in section~\ref{sec:sixpack}}
\label{fig:cm}
\end{figure*}

A modified plot of $c_i$ and $m$ at the three different signal-to-noise levels with all methods participating in GREAT08\footnote{cf. \citet{great08results} for explanations of the methods' acronyms} and including neural network blind estimates both with and without circularization is shown in Figure~\ref{fig:cm}.

The fact that the $m$ and $c$ found for blind data are consistent with the $m$ and $c$ found on training data is additional evidence that the networks we use are not overfitted to the training data. In 22 out of the 24 sets, components and bias parameters, linear bias corresponds within 1$\sigma$, in the other two cases within 2$\sigma$ of the bias measurement uncertainty.

\begin{deluxetable}{lll|rr|rr|rr|rrrr}
\tabletypesize{\scriptsize}
\rotate
\tablecaption{Performance and bias of different shear estimation methods on non-circularized samples\label{tbl:bias}}
\tablewidth{0pt}
\tablehead{
\colhead{method} & \colhead{sample\tablenotemark{a}} & \colhead{comp.} & & \colhead{$c$\tablenotemark{b}} & & \colhead{$m$\tablenotemark{b}} & & \colhead{$Q_1$\tablenotemark{c}} & \colhead{$Q_6$\tablenotemark{c}} & \colhead{$\sigma$\tablenotemark{d}} & \colhead{$b\times10^3$\tablenotemark{d}} \\
 & & & blind\tablenotemark{e} & all\tablenotemark{f} & blind\tablenotemark{e} & all\tablenotemark{f} & blind\tablenotemark{e} & all/train\tablenotemark{g} & & & &
}
\startdata
KSB$_S$    & 1 & 1 & \multicolumn{2}{|r|}{$2.95\pm0.06\times10^{-3}$} & \multicolumn{2}{|r|}{$-3.3\pm0.2\times10^{-2}$} &\multicolumn{2}{|r|}{$7.4\pm0.3$} & $10.0\pm0.9$ & 0.20 & 3.06 \\
           &   & 2 & \multicolumn{2}{|r|}{$5\pm6\times10^{-5}$}       & \multicolumn{2}{|r|}{$-1.4\pm0.3\times10^{-2}$} &\multicolumn{2}{|r|}{$21.5\pm0.8$} & $81\pm7$ & 0.20 & 0.74 \\
\tableline
KSB$_S$ aff\tablenotemark{h}& 1 & 1 & \multicolumn{2}{|r|}{$0\pm6\times10^{-5}$} & \multicolumn{2}{|r|}{$0\pm2\times10^{-3}$} & \multicolumn{2}{|r|}{$18.0\pm0.7$} & $47\pm4$ & 0.20 & 1.20 \\
           &   & 2 & \multicolumn{2}{|r|}{$0\pm6\times10^{-5}$} & \multicolumn{2}{|r|}{$0\pm3\times10^{-3}$} & \multicolumn{2}{|r|}{$21.5\pm0.8$}& $82\pm7$ & 0.20 & 0.73 \\
\tableline
KSB$_S$+NN\tablenotemark{i}& 1 & 1 & $0\pm2\times10^{-4}$ & $5\pm5\times10^{-5}$ & $-2\pm1\times10^{-2}$ & $-7\pm2\times10^{-3}$ & $30\pm6$ & $26.2\pm1.0$ & $133\pm12$ & 0.19 & 0.37 \\ 
           &   & 2 & $0\pm2\times10^{-4}$ & $-1\pm5\times10^{-5}$ & $-2\pm2\times10^{-2}$ & $-1.2\pm0.3\times10^{-2}$ & $31\pm6$ & $27.1\pm1.0$ & $141\pm13$ & 0.19 & 0.35 \\ 
\tableline
\tableline
KSB$_S$    & 2 & 1 & \multicolumn{2}{|r|}{$3.46\pm0.04\times10^{-3}$} & \multicolumn{2}{|r|}{$2.6\pm0.2\times10^{-2}$} & \multicolumn{2}{|r|}{$7.3\pm0.6$} & $8\pm2$ & 0.07 & 3.64 \\
           &   & 2 & \multicolumn{2}{|r|}{$-3.0\pm0.4\times10^{-4}$} & \multicolumn{2}{|r|}{$3.8\pm0.2\times10^{-2}$}  & \multicolumn{2}{|r|}{$95\pm8$} & $160\pm30$ & 0.07 & 0.74 \\
\tableline
KSB$_S$ aff\tablenotemark{h}& 2 & 1 & \multicolumn{2}{|r|}{$0\pm4\times10^{-5}$} & \multicolumn{2}{|r|}{$0\pm2\times10^{-3}$} & \multicolumn{2}{|r|}{$187\pm15$} & $1000\pm200$ & 0.07 & 0.13 \\
           &   & 2 & \multicolumn{2}{|r|}{$0\pm4\times10^{-5}$} & \multicolumn{2}{|r|}{$0\pm2\times10^{-3}$} & \multicolumn{2}{|r|}{$210\pm20$} & $1300\pm300$ & 0.07 & 0.00 \\
\tableline
KSB$_S$+NN\tablenotemark{i}& 2 & 1 & $0\pm1.5\times10^{-4}$ & $2\pm4\times10^{-5}$ & $2\pm5\times10^{-3}$ & $-1\pm2\times10^{-3}$ & $200\pm60$ & $220\pm20$ & $1000\pm200$ & 0.07 & 0.16 \\ 
           &   & 2 & $0\pm2\times10^{-4}$ & $0\pm4\times10^{-5}$ & $-3\pm9\times10^{-3}$ & $-2\pm2\times10^{-3}$ & $160\pm50$ & $260\pm20$ & $1900\pm400$ & 0.06 & 0.00 \\ 
\tableline
\tableline
KSB$_S$    & 3 & 1 & \multicolumn{2}{|r|}{$8\pm2\times10^{-4}$} & \multicolumn{2}{|r|}{$-1.83\pm0.07\times10^{-1}$} & \multicolumn{2}{|r|}{$3.6\pm0.3$} & $5.0\pm1.0$ & 0.30 & 4.32 \\
           &   & 2 & \multicolumn{2}{|r|}{$5\pm2\times10^{-4}$} & \multicolumn{2}{|r|}{$-1.49\pm0.10\times10^{-1}$} & \multicolumn{2}{|r|}{$5.4\pm0.4$} & $10\pm2$ & 0.32 & 2.87 \\
\tableline
KSB$_S$ aff\tablenotemark{h}& 3 & 1 & \multicolumn{2}{|r|}{$0\pm2\times10^{-4}$} & \multicolumn{2}{|r|}{$0\pm7\times10^{-3}$} & \multicolumn{2}{|r|}{$7.2\pm0.6$} & $45\pm9$ & 0.37 & 0.00 \\
           &   & 2 & \multicolumn{2}{|r|}{$0\pm2\times10^{-4}$} & \multicolumn{2}{|r|}{$0\pm1\times10^{-2}$} & \multicolumn{2}{|r|}{$6.9\pm0.6$} & $36\pm7$ & 0.38 & 0.63 \\
\tableline
KSB$_S$+NN\tablenotemark{i}& 3 & 1 & $6\pm7\times10^{-4}$ & $3\pm2\times10^{-4}$ & $-4\pm3\times10^{-2}$ & $-2.4\pm0.8\times10^{-2}$ & $7\pm2$ & $9.7\pm0.9$ & $45\pm9$ & 0.31 & 0.79 \\ 
            &   & 2 & $-4\pm5\times10^{-4}$ & $-2\pm2\times10^{-4}$ & $-4\pm3\times10^{-2}$ & $-4\pm1\times10^{-2}$ & $12\pm3$ & $9.3\pm0.8$ & $41\pm8$ & 0.32 & 0.88 \\ 
\enddata
\tablenotetext{a}{cf. Table~\ref{tbl:samples} for a description of the individual samples; sample 1 is the largest one with medium signal-to-noise and inhomogeneous galaxy properties and thus the most realistic one}
\tablenotetext{b}{additive and multiplicative bias for the individual component, as fitted in section~\ref{sec:linbias}}
\tablenotetext{c}{$Q_6$ is the quality parameter as used by GREAT08, averaging residuals over 'six-packs' of 60000 galaxies from the complete sample, $Q_1$ averages residuals over single sets of 10000 galaxies only; cf. eqn.~\ref{eqn:q}. $Q_6/Q_1\approx 6$ for unbiased methods, $Q_6/Q_1\approx 1$ where bias strongly dominates the noise at this sample size; cf. section~\ref{sec:sixpack}}
\tablenotetext{d}{estimates of the rms of individual galaxy scatter $\sigma$ and bias $b$, averaging over the whole sample}
\tablenotetext{e}{quantity measured for KSB$_S$+NN on the blind data only}
\tablenotetext{f}{quantity measured on the complete sample, using training and blind data for KSB$_S$+NN}
\tablenotetext{g}{quantity measured on the complete sample for KSB$_S$ and KSB$_S$ aff, on training data only for KSB$_S$+NN}
\tablenotetext{h}{these are KSB$_S$ outputs after an affine transformation with the $m_i$ and $c_i$ for the respective sample, such that the multiplicative and additive bias as fitted in section~\ref{sec:linbias} disappear; cf. eqn.~\ref{eqn:aff} and section~\ref{sec:sixpack}}
\tablenotetext{i}{these are the neural network corrected KSB$_S$ results}
\end{deluxetable}

\begin{deluxetable}{lll|rr|rr|rr|rrrr}
\tabletypesize{\scriptsize}
\rotate
\tablecaption{Performance and bias of different shear estimation methods on circularized samples\label{tbl:biasc}}
\tablewidth{0pt}
\tablehead{
\colhead{method} & \colhead{sample\tablenotemark{a}} & \colhead{comp.} & & \colhead{$c$\tablenotemark{b}} & & \colhead{$m$\tablenotemark{b}} & & \colhead{$Q_1$\tablenotemark{c}} & \colhead{$Q_6$\tablenotemark{c}} & \colhead{$\sigma$\tablenotemark{d}} & \colhead{$b\times10^3$\tablenotemark{d}} \\
 & & & blind\tablenotemark{e} & all\tablenotemark{f} & blind\tablenotemark{e} & all\tablenotemark{f} & blind\tablenotemark{e} & all/train\tablenotemark{g} & & & &
}
\startdata
KSB$_S$    & 1c & 1 & \multicolumn{2}{|r|}{$-2.8\pm0.4\times10^{-4}$}   & \multicolumn{2}{|r|}{$0\pm2\times10^{-3}$}     & \multicolumn{2}{|r|}{$29.3\pm0.9$} & $84\pm4$ & 0.16 & 0.87 \\
           &    & 2 & \multicolumn{2}{|r|}{$-1.68\pm0.04\times10^{-3}$} & \multicolumn{2}{|r|}{$1.0\pm0.2\times10^{-2}$} & \multicolumn{2}{|r|}{$15.9\pm0.5$} & $24.0\pm1.3$ & 0.16 & 1.93 \\
\tableline
KSB$_S$ aff\tablenotemark{h}& 1c & 1 & \multicolumn{2}{|r|}{$0\pm4\times10^{-5}$} & \multicolumn{2}{|r|}{$0\pm2\times10^{-3}$} & \multicolumn{2}{|r|}{$30.0\pm0.9$} & $89\pm5$ & 0.16 & 0.82 \\
           &    & 2 & \multicolumn{2}{|r|}{$0\pm4\times10^{-5}$} & \multicolumn{2}{|r|}{$0\pm2\times10^{-3}$} & \multicolumn{2}{|r|}{$29.5\pm0.9$} & $77\pm5$ & 0.16 & 0.95 \\
\tableline
KSB$_S$+NN\tablenotemark{i}& 1c & 1 & $-8\pm5\times10^{-5}$ & $-3\pm4\times10^{-5}$ & $-5\pm2\times10^{-3}$ & $-5.2\pm1.5\times10^{-3}$ & $37.4\pm1.5$ & $40\pm2$ & $200\pm20$\tablenotemark{j} & 0.16 & 0.22 \\ 
           &   & 2 & $-1.4\pm0.4\times10^{-4}$ & $-7\pm3\times10^{-5}$ & $-1.6\pm0.3\times10^{-2}$ & $-1.3\pm0.2\times10^{-2}$ & $41\pm2$ & $46\pm2$ & $180\pm20$\tablenotemark{j} & 0.15 & 0.44 \\ 
\tableline
\tableline
KSB$_S$    & 2c  & 1 & \multicolumn{2}{|r|}{$9\pm5\times10^{-5}$} & \multicolumn{2}{|r|}{$2.5\pm0.2\times10^{-2}$} & \multicolumn{2}{|r|}{$100\pm8$} & $180\pm40$ & 0.07 & 0.67 \\ 
           &     & 2 & \multicolumn{2}{|r|}{$-1.94\pm0.04\times10^{-3}$} & \multicolumn{2}{|r|}{$3.6\pm0.2\times10^{-2}$} & \multicolumn{2}{|r|}{$21\pm2$} & $24\pm5$ & 0.07 & 2.03 \\
\tableline
KSB$_S$ aff\tablenotemark{h}& 2c & 1 & \multicolumn{2}{|r|}{$0\pm5\times10^{-5}$} & \multicolumn{2}{|r|}{$0\pm2\times10^{-3}$} & \multicolumn{2}{|r|}{$172\pm14$} & $670\pm130$ & 0.07 & 0.25 \\ 
           &     & 2 & \multicolumn{2}{|r|}{$0\pm4\times10^{-5}$} & \multicolumn{2}{|r|}{$0\pm2\times10^{-3}$} & \multicolumn{2}{|r|}{$210\pm20$} & $1300\pm300$ & 0.07 & 0.00 \\
\tableline
KSB$_S$+NN\tablenotemark{i}& 2c & 1 & $0.3\pm1.2\times10^{-4}$ & $0\pm4\times10^{-5}$ & $2\pm5\times10^{-3}$ & $-1\pm2\times10^{-3}$ & $230\pm70$ & $230\pm20$ & $1000\pm200$ & 0.06 & 0.18 \\ 
           &        & 2 & $-1\pm2\times10^{-4}$ & $-2\pm4\times10^{-5}$ & $-6\pm8\times10^{-3}$ & $-2\pm2\times10^{-3}$ & $160\pm50$ & $260\pm20$ & $1700\pm300$ & 0.06 & 0.00 \\ 
\tableline
\tableline
KSB$_S$    & 3c & 1 & \multicolumn{2}{|r|}{$-2.8\pm0.2\times10^{-3}$} & \multicolumn{2}{|r|}{$-1.35\pm0.07\times10^{-1}$} & \multicolumn{2}{|r|}{$3.2\pm0.3$} & $4.1\pm0.8$ & 0.29 & 4.81 \\
           &    & 2 & \multicolumn{2}{|r|}{$-8\pm2\times10^{-4}$} & \multicolumn{2}{|r|}{$-1.08\pm0.09\times10^{-1}$} & \multicolumn{2}{|r|}{$7.5\pm0.6$} & $16\pm3$ & 0.29 & 2.24 \\
\tableline
KSB$_S$ aff\tablenotemark{h}& 3c & 1 & \multicolumn{2}{|r|}{$0\pm2\times10^{-4}$} & \multicolumn{2}{|r|}{$0\pm7\times10^{-3}$} & \multicolumn{2}{|r|}{$8.5\pm0.7$} & $40\pm8$ & 0.33 & 0.82 \\
           &    & 2 & \multicolumn{2}{|r|}{$0\pm2\times10^{-4}$} & \multicolumn{2}{|r|}{$0\pm9\times10^{-3}$} & \multicolumn{2}{|r|}{$9.2\pm0.7$} & $47\pm9$ & 0.32 & 0.61 \\
\tableline
KSB$_S$+NN\tablenotemark{i} & 3c & 1 &  $5\pm6\times10^{-4}$ & $2\pm2\times10^{-4}$ & $-9\pm3\times10^{-2}$ & $-3\pm1\times10^{-2}$ & $7\pm2$ & $11.0\pm0.9$ & $45\pm9$ & 0.29 & 0.91 \\ 
           &    & 2 & $-3\pm5\times10^{-4}$ & $-2\pm2\times10^{-4}$ & $1\pm3\times10^{-2}$ & $-2\pm1\times10^{-2}$ & $13\pm4$ & $12.0\pm0.9$ & $55\pm11$ & 0.28 & 0.69 \\ 
\enddata
\tablenotetext{a}{cf. Table~\ref{tbl:samples} for a description of the individual samples; sample 1c is the largest one with medium signal-to-noise, inhomogeneous galaxy properties and different PSFs and thus the most realistic one}
\tablenotetext{b}{additive and multiplicative bias for the individual component, as fitted in section~\ref{sec:linbias}}
\tablenotetext{c}{$Q_6$ is the quality parameter as used by GREAT08, averaging residuals over 'six-packs' of 60000 galaxies from the complete sample, $Q_1$ averages residuals over single sets of 10000 galaxies only; cf. eqn.~\ref{eqn:q}. $Q_6/Q_1\approx 6$ for unbiased methods, $Q_6/Q_1\approx 1$ where bias strongly dominates the noise at this sample size; cf. section~\ref{sec:sixpack}}
\tablenotetext{d}{estimates of the rms of individual galaxy scatter $\sigma$ and bias $b$, averaging over the whole sample}
\tablenotetext{e}{quantity measured for KSB$_S$+NN on the blind data only}
\tablenotetext{f}{quantity measured on the complete sample, using training and blind data for KSB$_S$+NN}
\tablenotetext{g}{quantity measured on the complete sample for KSB$_S$ and KSB$_S$ aff, on training data only for KSB$_S$+NN}
\tablenotetext{h}{these are KSB$_S$ outputs after an affine transformation with the $m_i$ and $c_i$ for the respective sample, such that the multiplicative and additive bias as fitted in section~\ref{sec:linbias} disappear; cf. eqn.~\ref{eqn:aff} and section~\ref{sec:sixpack}}
\tablenotetext{i}{these are the neural network corrected KSB$_S$ results}
\tablenotetext{j}{$Q_6$ as found on the blind data only in the case of sample 1c}
\end{deluxetable}

\subsection{Bias analysis II: 'six-pack' effect}
\label{sec:sixpack}

The bias of a method likely differs depending on properties such as signal-to-noise ratio, PSF, galaxy size and profiles. Therefore in the case of inhomogeneous samples like sample 1 and 1c, the $c_i$ and $m$ we find for the whole sample are not necessarily equal to the true additive and multiplicative biases of each homogeneous subsample. For this reason, a method simply calibrated for $c_i=m\approx0$ by an affine transformation for the whole inhomogeneous sample is potentially still biased on each of the homogeneous subsamples and consequently for any real-world application. Thus in order to correctly analyze the bias by means of fitting multiplicative and additive biases, each sample would have to be splitted into homogeneous subsamples first. We develop another scheme of analyzing bias here which can be applied to homogeneous and inhomogeneous samples alike.

The GREAT08 data sets which we used for training the networks are made up of files containing 10,000 galaxies each, six of which again share the exact same shear values, observing conditions in terms of PSF, signal-to-noise ratio and galaxy properties. This is a very favorable setting for bias analysis, as the composition of errors from bias and scatter can be estimated by comparing the accuracy of shear estimates on single sets and on six times larger overall sets.

When measuring the shear of a very large homogeneous set $j$ of $n\rightarrow\infty$ galaxies with true shear $(g_1^j,g_2^j)$, the only residual for component $i$ will be the bias $b_i^j$. For a linear bias $m_i^j$ and $c_i^j$, we would find $b_i^j=m_i^j\cdot g_i^j + c_i^j$. In the limit of infinitely large sets the squared error of the shear estimate $\bar{e}_i^j$ will be
\begin{equation}
Q^{-1}=(\bar{e}_i^j-g_i^j)^2 \stackrel{n\rightarrow\infty}{\rightarrow}(b_i^j)^2 \; .
\end{equation}

For smaller $n$, the scatter $\sigma_j$ of the individual galaxy measurement in that set will add to the errors, leading to
\begin{equation}
Q^{-1}=(\bar{e}_i^j-g_i^j)^2 =  (b_i^j)^2 + \frac{\sigma^2}{n}\; .
\label{eqn:sixplot}
\end{equation}
This bias and the scatter will of course depend of the properties of the particular set $j$. For the following analysis, we are interested only in a decomposition of our total mean squared error into bias and scatter. We will thus calculate $\langle(\bar{e}_i^j-g_i^j)^2\rangle_j$ at two different sample sizes $n=(10,000, 60,000)$ to find the root mean square of the bias $b_i:=\sqrt{\langle(b_i^j)^2\rangle_j}$ and the scatter $\sigma:=\sqrt{\langle(\sigma^j)^2\rangle_j}$.

\begin{figure}[h!]
\centering
\includegraphics[width=0.4\textwidth]{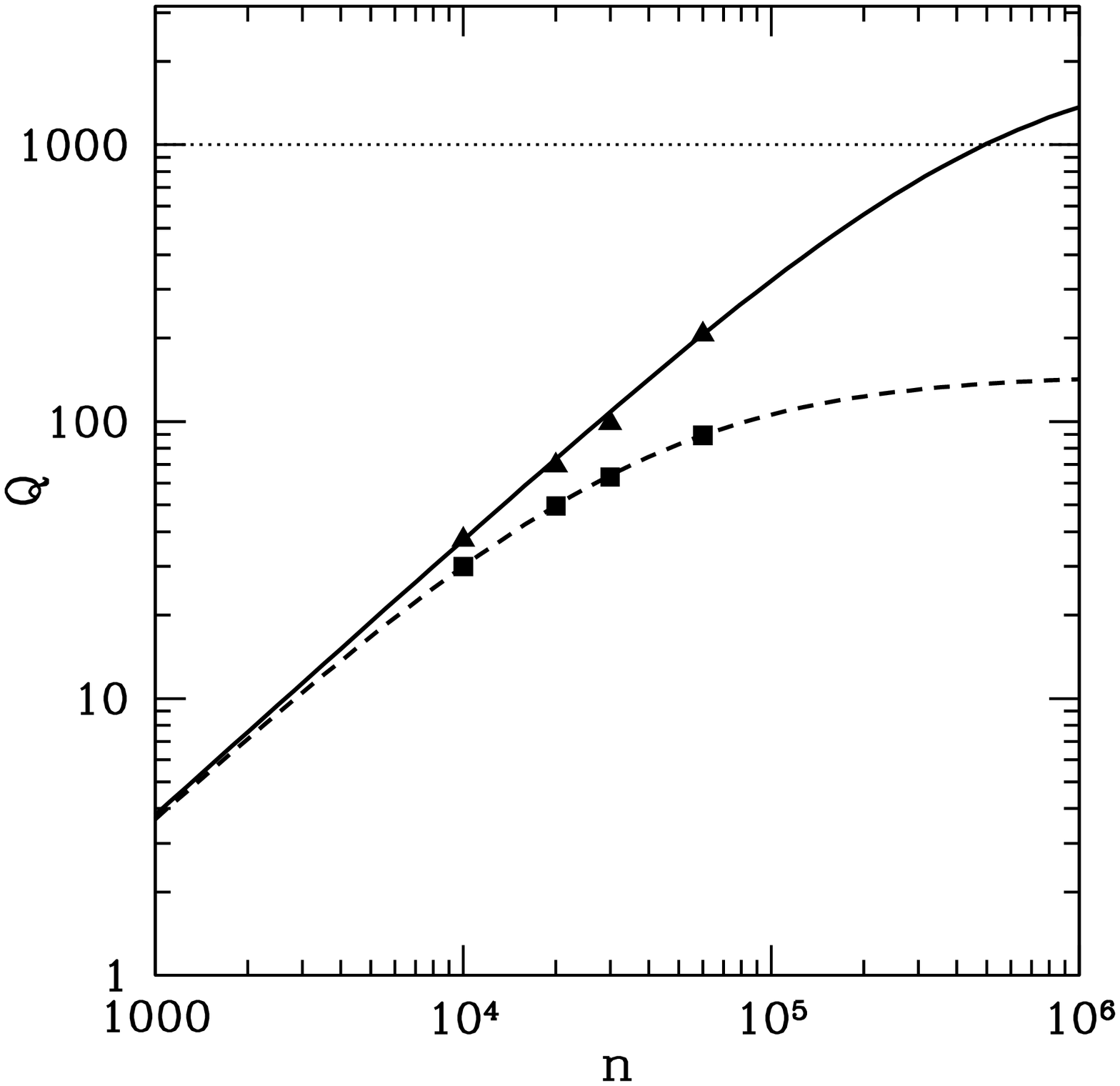}
\caption{$Q$ at different sample sizes. Plotted are measured $Q$ values for $n=10,000, 20,000, 30,000, 60,000$ galaxies on component 1 of sample 1c for the neural network blind data estimate (triangles / solid line) and the affine transformation of KSB outputs with $m=0$, $c=0$  (squares / dashed line). The curves show the $Q(n)$ from eqn.~\ref{eqn:sixplot} for the bias and scatter as found in Table~\ref{tbl:biasc} for the two methods.}
\label{fig:sixplot}
\end{figure}

We perform the analysis on the six samples. We calculate $Q_6$, similar to equation~\ref{eqn:q} yet averaging residuals over the six sets before taking the square. This is what the leaderboard for the GREAT08 challenge used for ranking the submissions. Note that for pure scatter we would find $Q_6 / Q_1 = 6$, whereas for pure bias $Q_6 / Q_1 = 1$.  This favors methods with low bias which perform consistently well for different conditions and galaxy parameters. For large future surveys it is of particular importance to achieve a bias which is low even compared to the smaller scatter of the large sample sizes (i.e. $Q_n / Q_1 \approx n$), regardless of the individual galaxy properties or observing conditions. Except for sample 1c where the number of blind sets available is large enough and completeness of the sets of 60000 galaxies has been conserved when separating the blind sets, we have to use the complete sample of galaxy sets (i.e. both the sets used for training and for blind testing) for calculating $Q_6$. Where both $Q_1$ and the linear bias are consistent between the training and blind sets, this is not expected to give different results than a test on purely blind data.

We also compare the network results to the results of KSB$_S$ output scaled with the parameters $m_i$ and $c_i$ found for KSB$_S$ on each of the samples in section \ref{sec:linbias}.  The result, defined as
\begin{equation}
e_i^{\mathrm{KSB\;aff.}}=\frac{e_i^{\mathrm{iso}}/0.91-c_i}{m_i+1}\; .
\label{eqn:aff}
\end{equation}
is denoted as KSB aff. in Figure~\ref{fig:cm} and Tables~\ref{tbl:bias} and \ref{tbl:biasc}. By definition, KSB aff. has $m=c_i=0$ on any of the samples. Note that signal-to-noise dependence of the bias is likely the strongest influence on bias of all the GREAT08 parameters.\footnote{cf. Figures C3 and C4 in \citet{great08results}} A correction of this dependence by splitting into different signal-to-noise subsamples with different affine scalings, as has been done in the case of $e_i^{\mathrm{KSB\;aff.}}$, is therefore the most promising method if one additional parameter is taken into account for bias correction. The dependence of the bias on the signal-to-noise ratio has in fact been taken into account empirically in \citet{schrabback}. The fact that the neural network estimate outperforms $e_i^{\mathrm{KSB\;aff.}}$ on the inhomogeneous samples 1 and 1c shows that the networks successfully take other dependences of the bias into account. For direct comparison, a plot of the quality parameter $Q$ for the network estimate and KSB aff. as found for component 1 at different set sizes $n$ on the circularized sample 1c is shown in Figure~\ref{fig:sixplot}. Drawn are measured $Q$ at sample sizes of $n=10000,20000,30000,60000$. Note that this corresponds very well with the curve for eqn.~\ref{eqn:sixplot} drawn in terms of $b$ and $\sigma$ as found in Table~\ref{tbl:biasc}. The larger residual bias of KSB aff. on the subsets of the inhomogeneous sample leads to a much smaller asymptotic $Q$ than for the neural network estimate, which is projected to reach $Q>1000$ at sufficiently large sample sizes.

\begin{figure}
\centering
\subfigure[non-circularized samples] {
\plotone{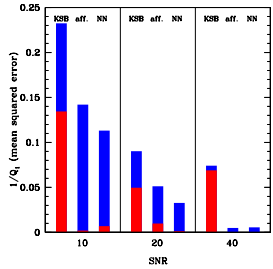}
}
\subfigure[circularized samples] {
\plotone{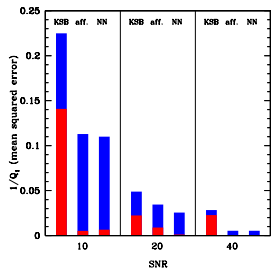}
}
\caption{Composition of mean squared errors (red: bias, blue: scatter) of the different methods (plain KSB$_S$: KSB; KSB$_S$ after affine transformation fitted to set multiplicative and additive bias to zero: aff.; KSB with neural network corrections: NN with total error as found for the blind sets) on samples with different signal-to-noise ratio. The middle blocks contains the medium signal-to-noise samples 1/1c with inhomogeneous galaxy properties most similar to real data.}
\label{fig:mse}
\end{figure}

\begin{figure}
\centering
\subfigure[non-circularized samples] {
\includegraphics[width=0.4\textwidth]{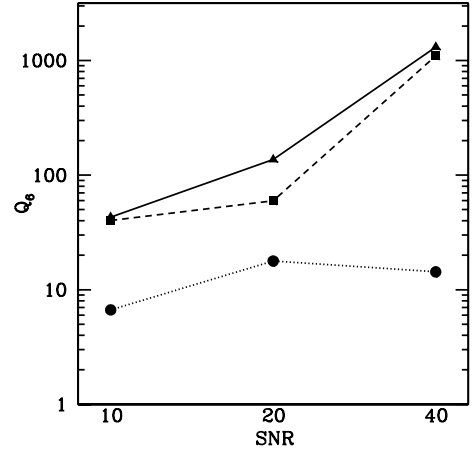}
}
\subfigure[circularized samples] {
\includegraphics[width=0.4\textwidth]{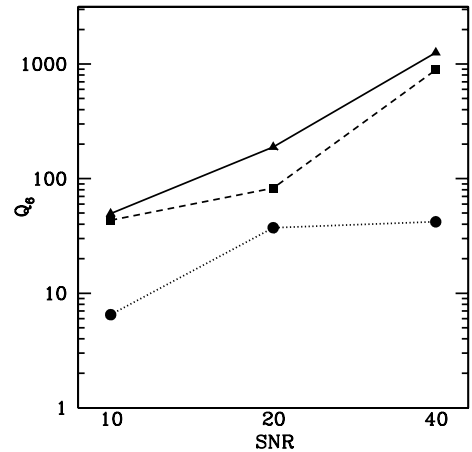}
}
\caption{$Q_6$ as a function of signal-to-noise ratio for the neural network estimate (solid line, triangles), KSB aff. (dashed line, squares) and plain KSB (dotted line, circles) plotted for non-circularized (upper panel) and non-circularized (lower panel) samples.}
\label{fig:q6}
\end{figure}

Complete results for plain KSB$_S$, KSB aff. and the output of best networks are shown in Tables~\ref{tbl:bias} and \ref{tbl:biasc}. A plot of the composition of squared errors is plotted in Figure~\ref{fig:mse}, while the GREAT08 quality parameter $Q_6$ as a function of signal-to-noise ratio is shown in Figure~\ref{fig:q6}.
The bias is predominant in plain KSB$_S$ outputs, especially for the first component. This is particularly harmful as sample size and signal strength increases. Despite lowered statistical uncertainty the strong bias leads to overall improvements being only slight.

Affine transformations according to a fit to the known true shears greatly reduce the bias on the homogeneous subsamples 2, 2c, 3 and 3c. This is not surprising, as data sets here only differ by the shear applied to them and the change in bias due to this can be accounted for. On samples with inhomogeneous galaxy properties such as samples 1 and 1c, however, bias cannot be removed by this and remains significant. Also, the remaining scatter due to noise and differences in bias depending on the individual galaxy properties within the sample cannot be decreased by the affine transformation.

Neural networks, on the contrary, greatly reduce the bias in any of the samples such that it does not dominate the statistical errors at this sample size. This works remarkably well even on the inhomogeneous samples. The observed reduction in scatter indicates that bias depending on the individual galaxy properties has been successfully reduced as well. For a bias reduction scheme to be applied to real data with diverse properties, this is of crucial importance. Therefore, as has been done with the neural networks, it can be seen that a linear bias of inhomogeneous samples close to zero should merely be achieved as a side effect of a proper overall calibration, which can be validated using other types of analyses as well.

\subsection{Circularization}
\label{sec:circ}
In order to compare the effects of PSF circularization to traditional $P^{\mathrm{sm}}$ anisotropy correction, we compare the results of sample 1, 2 and 3 to sample 1c, 2c and 3c. While these are similar in terms of signal-to-noise levels and galaxy properties,\footnote{Sample 1c contains 600 additional galaxies not in sample 1, which are of fiducial properties but convolved with PSF 2 or 3.} they differ in the method used for anisotropy correction. 

From the individual measurement $\sigma$ calculated in Tables~\ref{tbl:bias} and \ref{tbl:biasc} we find that scatter can be reduced by circularization, especially for the more noisy data sets. This can also be seen from comparing the left and right panels of Figure~\ref{fig:linbias} and is not surprising as the smear responsitivity tensor $P^{\mathrm{sm}}$ of the individual galaxy otherwise used for anisotropy correction certainly is influenced by the noise. The additional term $P^{\mathrm{sm}}\bm{p}$ in eqn.~\ref{eqn:ksb3} adds to the scatter. Because $\bm{p}=0$ for a circular PSF, this is not the case in the circularized samples.

Circularization, however, appears to add to or at least change the bias present in the KSB$_S$ output. On the medium to low signal-to-noise data, calibration by the neural networks is successful such that the overall result can be improved. On the contrary, a fitted affine transformation of KSB$_S$ output on sample 1c is still strongly biased.

For sufficiently noisy data, circularization can thus successfully be used as a method of reducing scatter. Plain KSB$_S$ outputs, however, even after traditional corrections, do not greatly profit from this as the residual bias dominates here. In order to benefit from the reduced scatter in circularized samples one has to combine circularization with a means of reducing bias, as has been done with the neural networks in this work.

\subsection{Analyzing network output}
\label{sec:percentile}
We continue to analyze the single galaxy network output as a function of KSB $\bm{e}^{\mathrm{iso}}$ for different subsamples of GREAT08.

\begin{figure*}
\centering
\subfigure[sample 1c (medium signal to noise), both components]{
\plotone{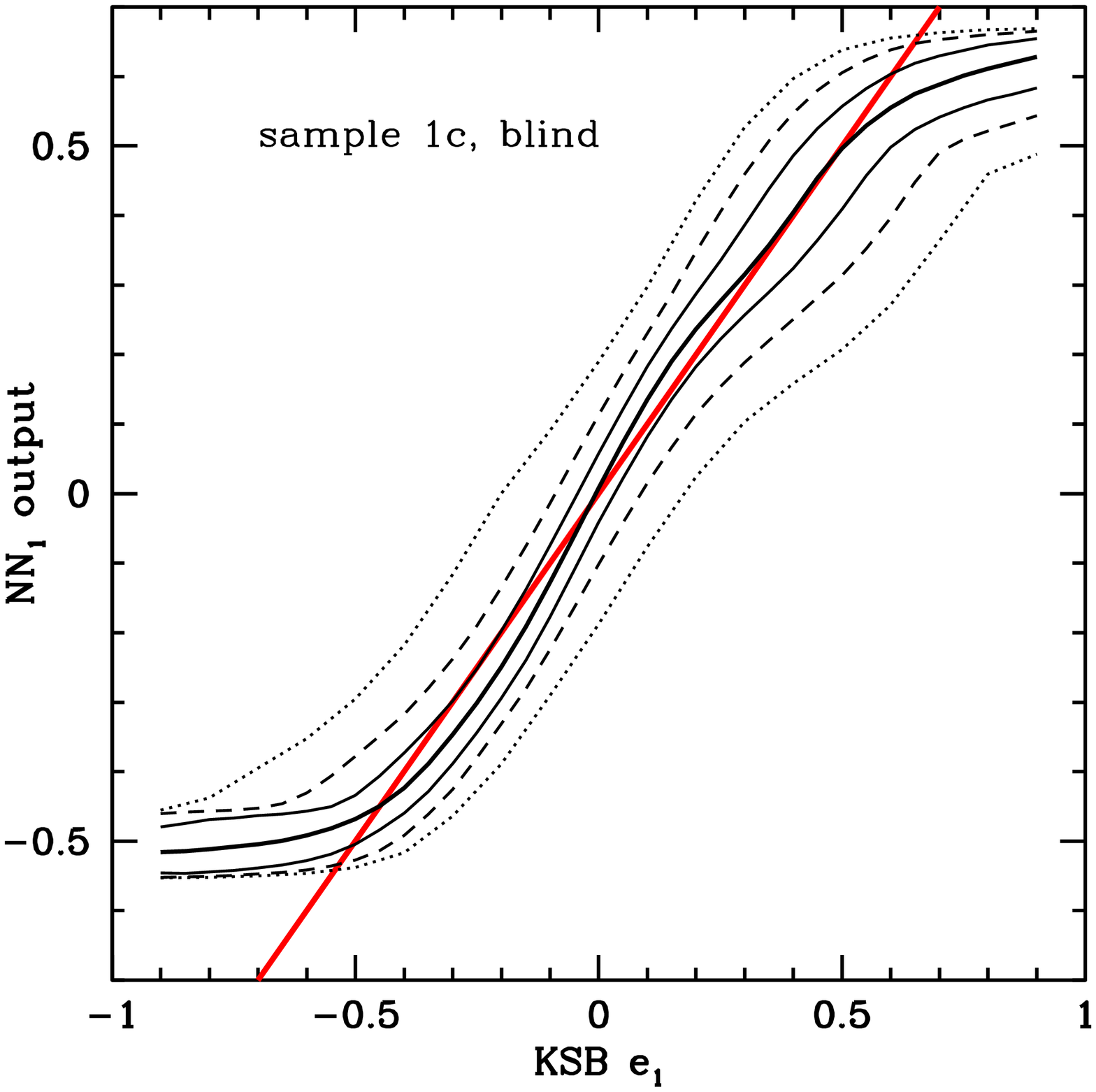}
\plotone{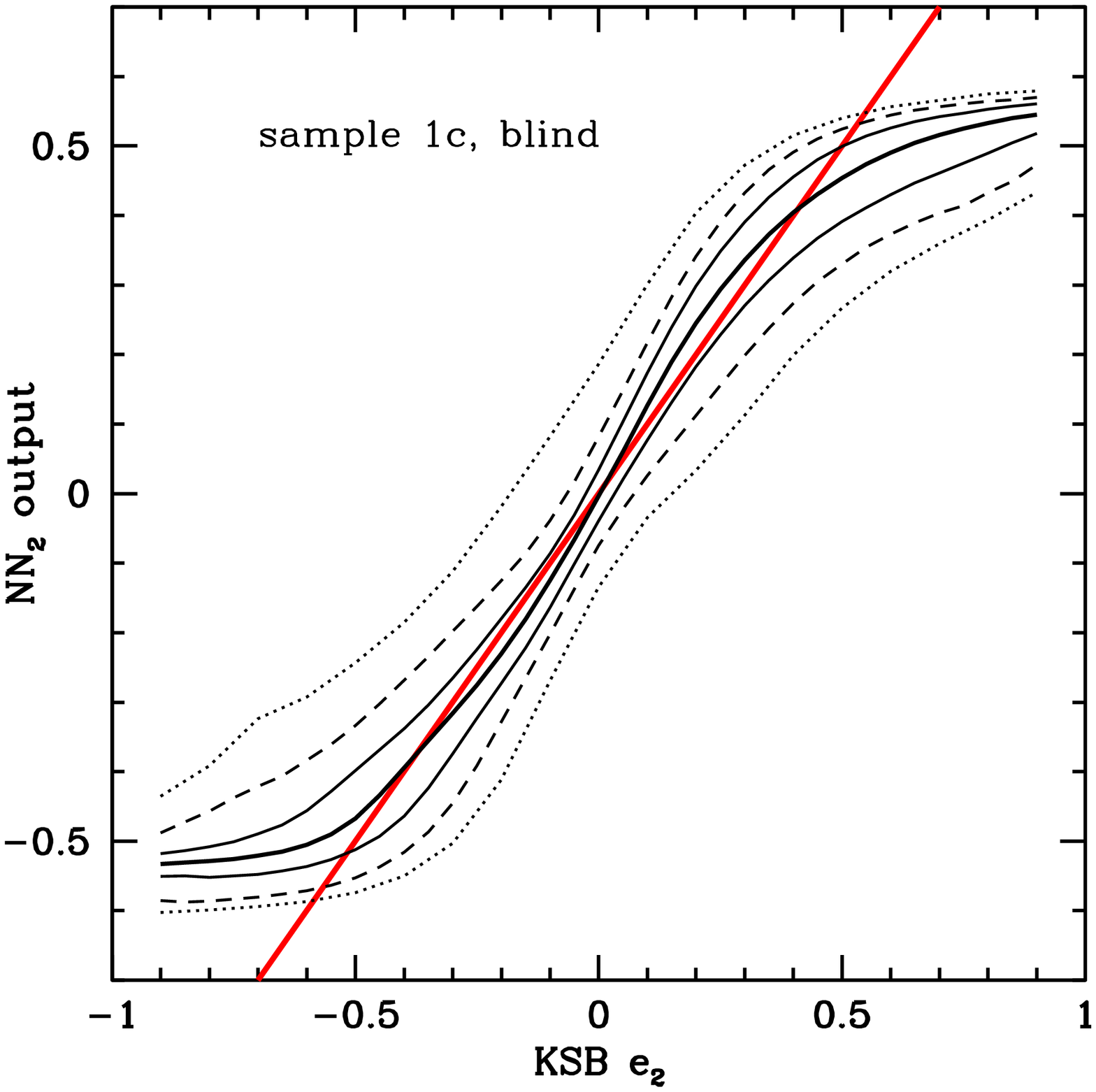}
}
\subfigure[sample 2c (high) and 3c (low signal to noise)]{
\plotone{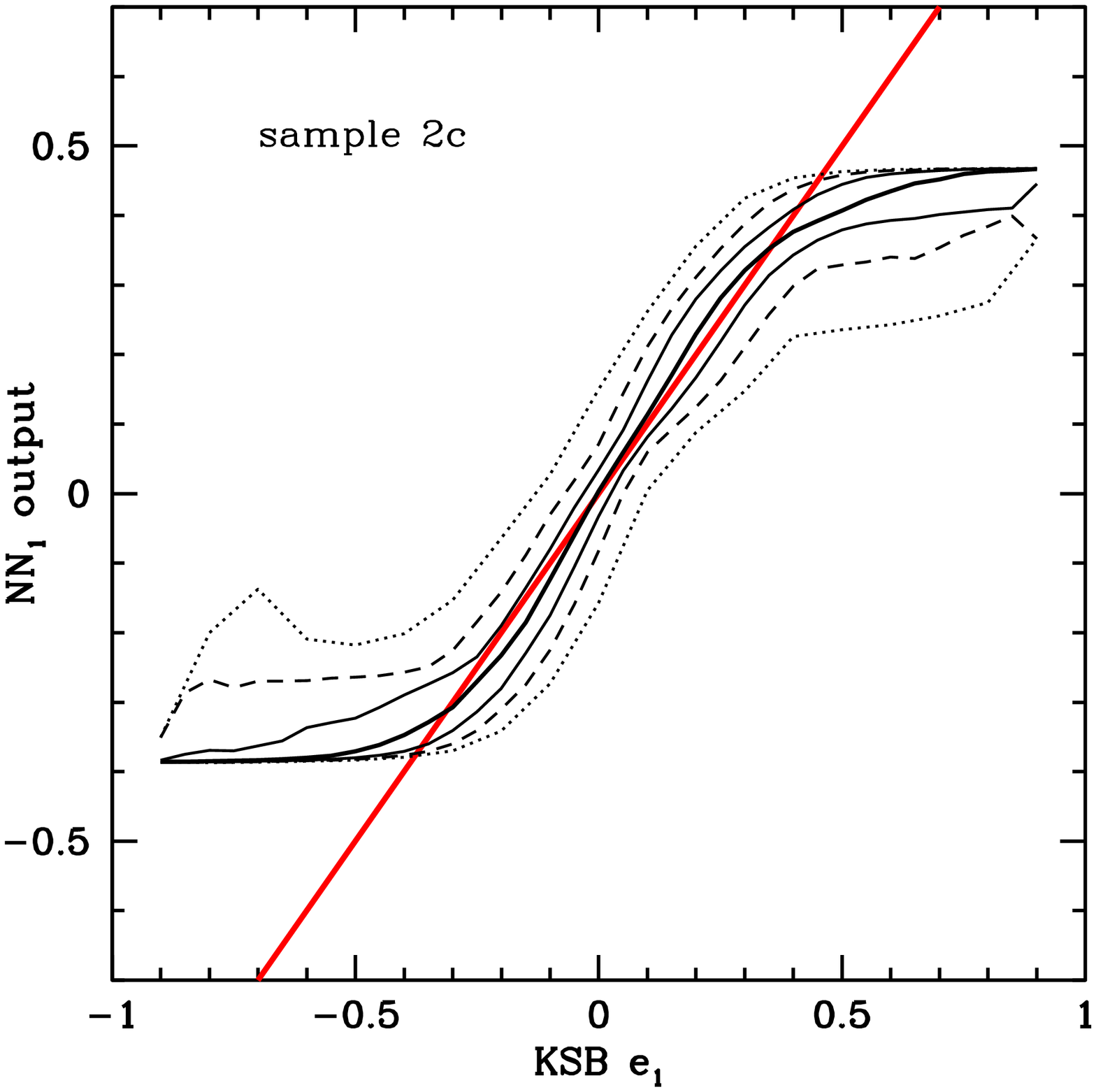}
\plotone{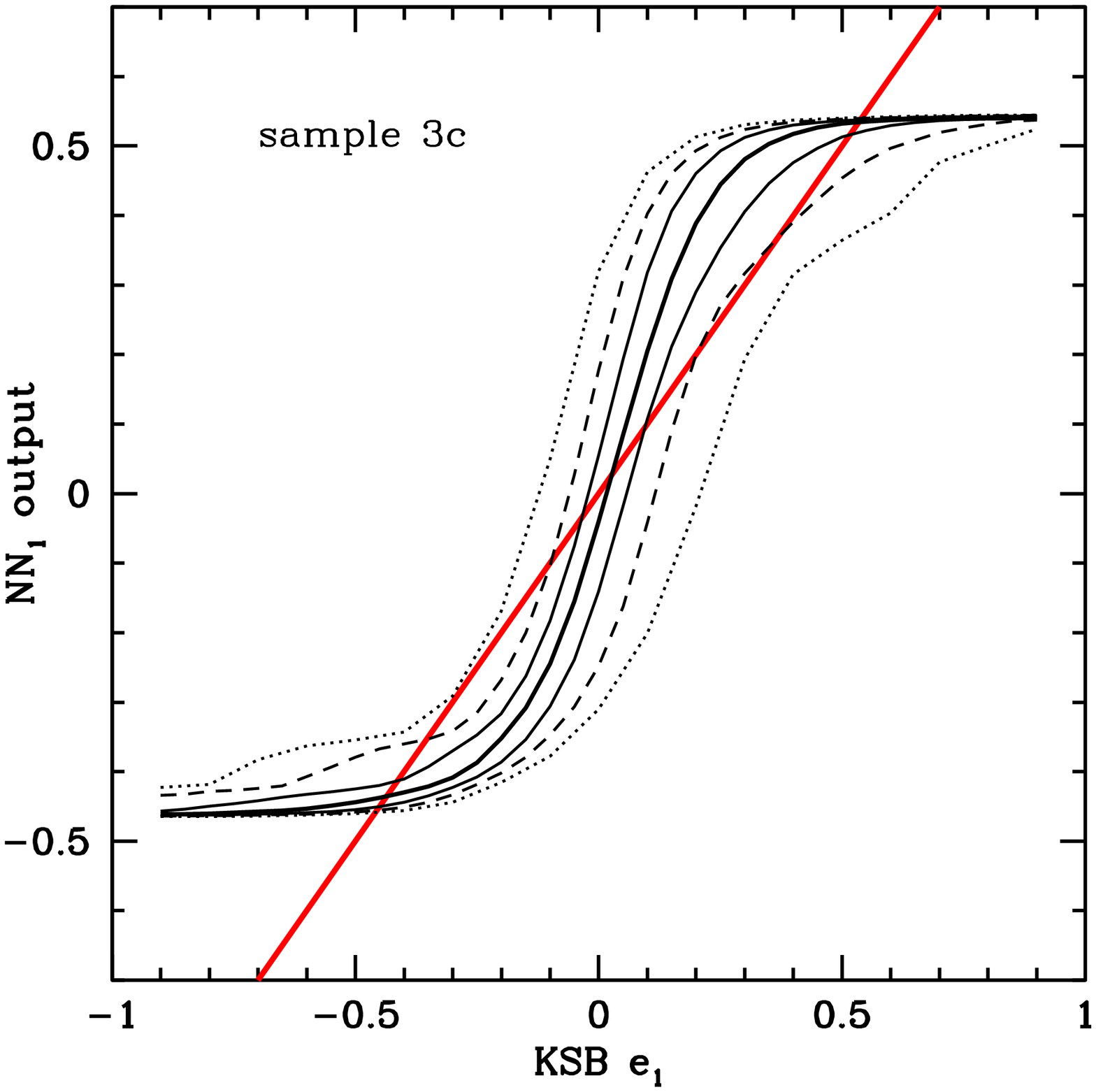}
}
\subfigure[branches from 1c with large and small galaxy FWHM]{
\plotone{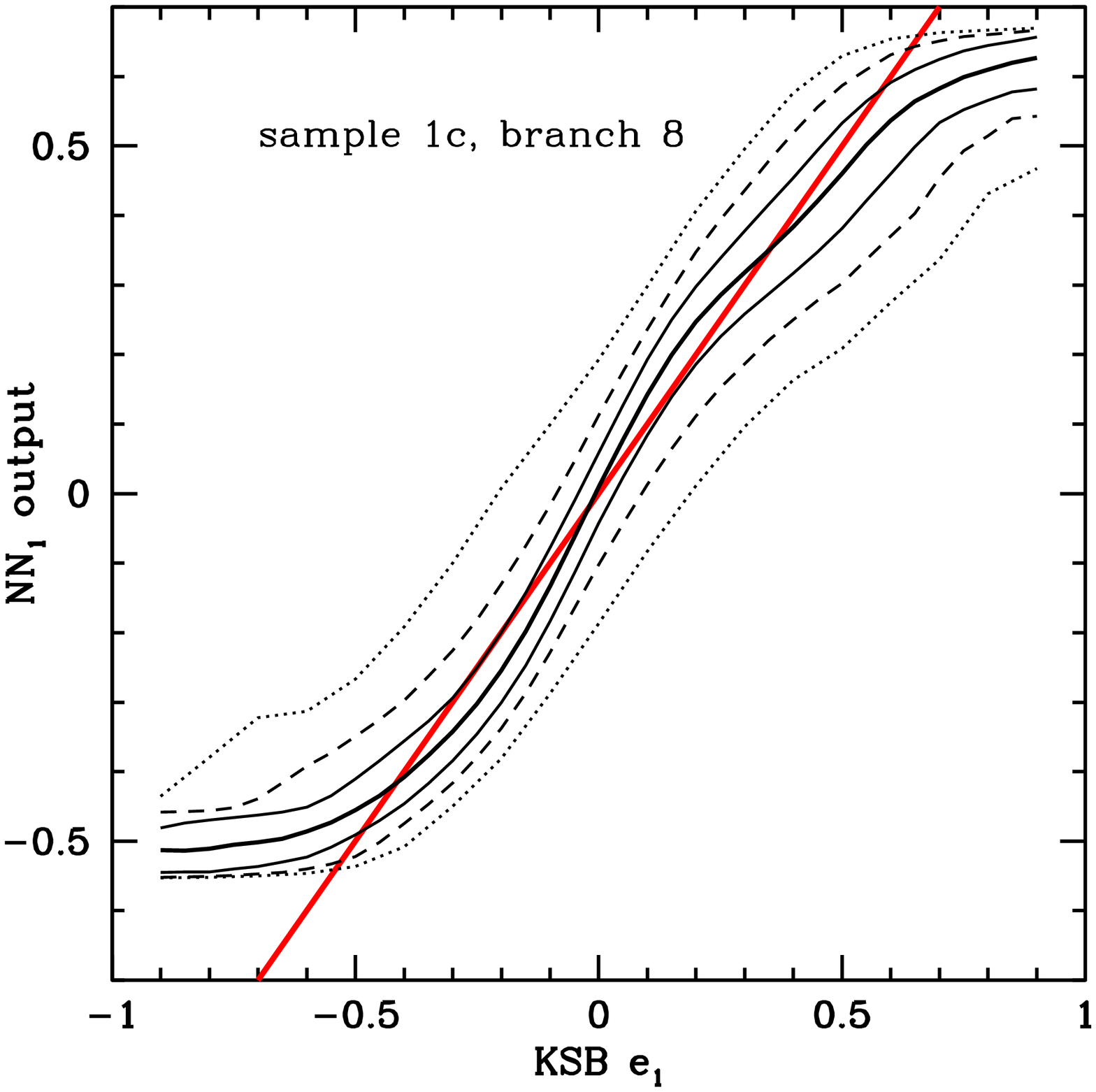}
\plotone{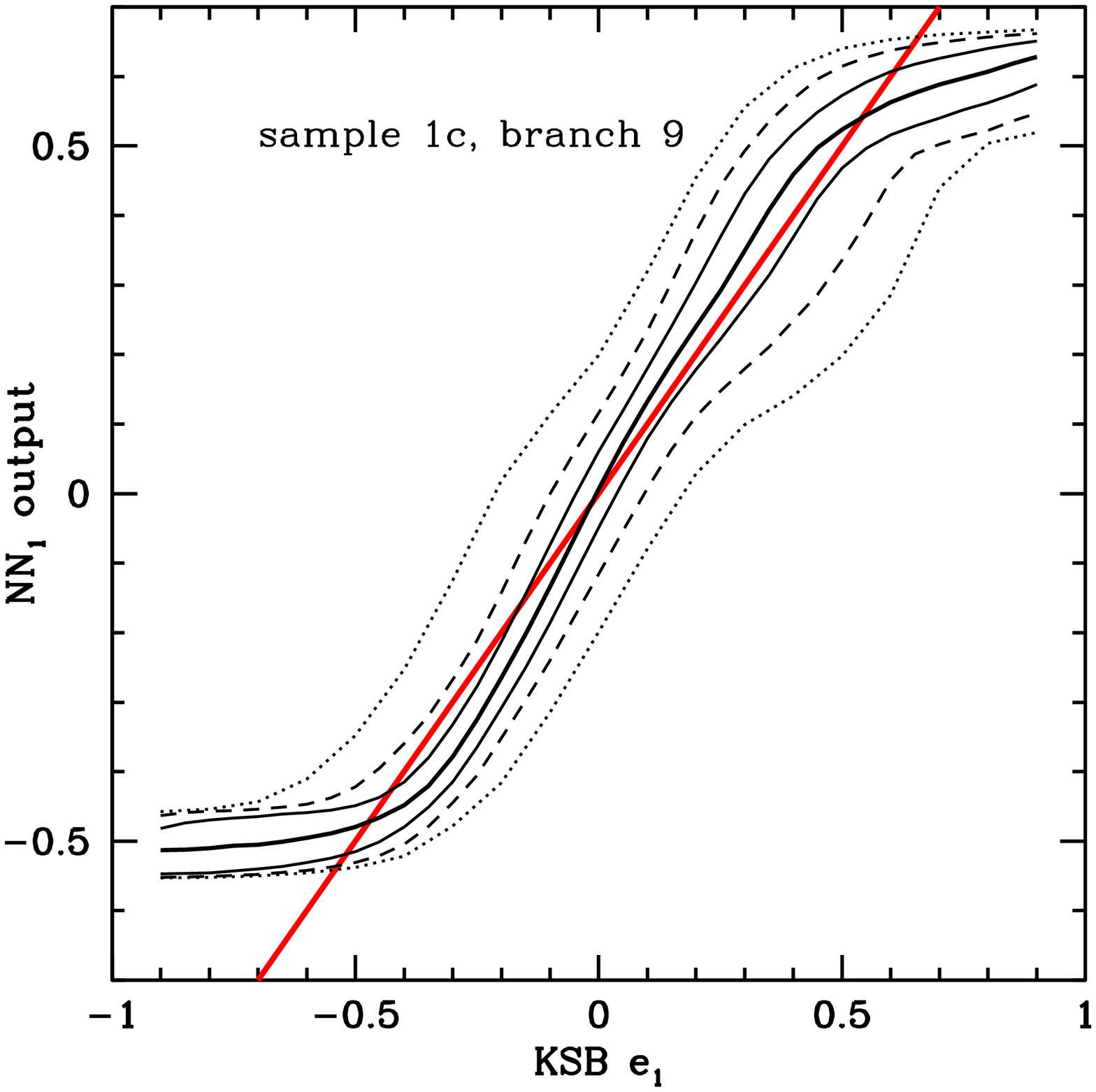}
}
\caption{Percentiles (50th corresponding to the median, 14th and 86th ($1\sigma$), 2.5th and 97.5th ($2\sigma$), 0.2th and 99.8th ($3\sigma$) of single galaxy network output as a function of KSB $\bm{e}^{\mathrm{iso}}$ for circularized samples. The red line is the identity. Both components are plotted for the overall sample 1c in the top panel, component 1 is plotted for non-fiducial signal-to-noise sets 2c and 3c in the middle panel and for non-fiducial galaxy FWHM in the lower panel.}
\label{fig:percentiles}
\end{figure*}

The network output for shear component $j$ is a nonlinear function $f_j(\bm{v}^i)$ of the input vector with KSB shear estimates and additional parameters $\bm{v}^i$ of each galaxy $i$. One may interpret this, although not unambiguously, as an additive bias correction $-c_j(\bm{v}^i)$ and a weighting and multiplicative bias correction $w_j(\bm{v}^i)$
\begin{equation}
w_j(\bm{v}^i):=\left(1+m_j(\bm{v}^i)\right)^{-1}\cdot\frac{\langle\sigma_j^2\rangle}{\sigma_j^2(\bm{v}^i)} \; ,
\end{equation}
where $\langle\sigma_j^2\rangle$ is the average variance of the measurements from the sample galaxy $i$ has been taken from, $\sigma_j^2(\bm{v}^i)$ the variance of the measurement of the particular galaxy and $m_j(\bm{v}^i)$ a galaxy set property dependent multiplicative bias. The network output can then be written as
\begin{equation}
f_j(\bm{v}^i)=w_j(\bm{v}^i)\cdot\left(e^{\mathrm{iso},i}_j-c_j(\bm{v}^i)\right) \; .
\end{equation}

Consider now a version of the galaxy rotated by 90$^{\circ}$ so as to take the true ellipticity $e_j$ to its negative $e_j'=-e_j$ and $\bm{v}^i\rightarrow\bm{v}^{i'}$. As the variance $\sigma_j^2(\bm{v}^i)$ and multiplicative bias $m_j(\bm{v}^i)$ should remain constant under such a transformation, an ideal network should use an unchanged $w_j(\bm{v}^{i'})=w_j(\bm{v}^i)$. The additive bias correction should be taken such that $e^{\mathrm{iso},i}_j-c_j(\bm{v}^i)=-\left(e^{\mathrm{iso},i'}_j - c_j(\bm{v}^{i'})\right)$. This results in a point symmetric distribution of neural network outputs as a function of $e^{\mathrm{iso}}_j$ with respect to a zero shifted by the additive bias. Differences in $c_j$ for different galaxies will cause asymmetries in the distributions, but for $|c_j|\ll1$ these will only be slight. We therefore expect a non-overfitted network to give an almost point symmetric output distribution.

For $\bm{e}^{\mathrm{iso}}$ bins we find percentiles of the network output corresponding to the median and 1-3$\sigma$ in the case of a normal distribution. Plots of the resulting percentile curves of the networks trained on circularized data are shown in Figure~\ref{fig:percentiles}. They show point symmetry, which is additional evidence that the networks we use are not overfitted. The general shape can be interpreted as a high weighting of relatively circular galaxies and a downweighting of more elliptical galaxies. Slight differences can also be seen, for instance, between the output for the subsamples of sample 1c with larger and smaller than fiducial galaxy FWHM on the left and right of the lower panel, respectively. None of the network corrections can be interpreted as a single affine transformation.

The network output, therefore, must not be interpreted as the \emph{true} ellipticity of the galaxies. It is merely a quantity that, if averaged arithmetically, gives a good ensemble shear estimate. For different applications, such as shear correlation measurements, different network inputs and cost functions can be used. Apart from defining a figure of merit and constructing the cost function such that the figure of merit is being maximized during training, one may in these cases make use of the rotational and permutational invariance of the expected output. 

\section{Conclusion}

We have presented a scheme of neural networks which is capable of reducing bias in KSB shear measurements to a level where it no longer inhibits the success of future surveys. Bias correction was most successful on the medium to high signal-to-noise data sets. This result might give hints as to the most promising setup of future pipelines.

We showed that circularization of the PSF reduces the scatter as compared to PSF anisotropy correction based on a single galaxy $P^{\mathrm{sm}}\bm{p}$ term. Therefore, circularization of varying PSFs in combination with neural networks seems to be very promising for shear measurement on real data.

Overall results in terms of shear measurement accuracy are very encouraging. By means of neural networks, it was possible to calibrate traditional KSB shape measurement to an accuracy competitive with the most successful methods and well above traditionally calibrated shape measurement approaches participating in the GREAT08 challenge. On real data KSB remains the method most commonly used, which makes this improvement extremely valuable. The neural network scheme presented in this paper is, however, also a general approach. It can be applied to any other shear measurement pipeline that is using single galaxy parameters to find true shears by an averaging procedure. We expect that neural networks are able to reduce bias in these methods as well.

The success of any shear measurement calibration scheme, including neural networks, depends on the availability of data with known shear similar to the data to be analyzed. In the case of the GREAT08 challenge, this was available from the simulations themselves. For the application on real data, it is necessary to simulate training data sets with known shear values from and similar to the real data. This can be done by either fitting galaxy models to the objects to be analyzed and simulating sheared data from the fitted profiles or by applying a finite-resolution shear operator \citep{fso} to the original image data itself.

\acknowledgements
\emph{This work was supported by the TR33 "The Dark Universe", the DFG 
Cluster of Excellence on the "Origin and Structure of the Universe" and 
the RTN-Network "DUEL"  (Dark Universe through Extragalactic Lensing) 
gravitational lensing. We thank the Bonn lensing group for an introduction to KSB and for providing us with some of their scripts. We are also grateful to A. Collister and O. Lahav for making available their neural network implementation ANNz which the scheme presented has been built upon.}

\appendix

\section{Appendix}
\subsection{Back-Propagation of Errors}
\label{sec:bp}
The learning procedure of a neural network consists in finding a set of weights for the connections between nodes of adjacent layers $w_{ij}$ such that some cost function $E$ of the network output becomes minimal. In principle, it is possible to achieve this using gradient descent, i.e. changing the weights in each step according to
\begin{equation}
\Delta w_{ij} = - \eta \frac{\partial E}{\partial w_{ij}} \; ,
\label{eqn:graddesc}
\end{equation}
with $\eta$ being a small, positive parameter. The updating of weights can be done on the basis of individual data sets or after a batch of data sets have been processed. However, an efficient method of calculating the required derivative $\frac{\partial E}{\partial w_{ij}}$ needs to be found.

For the output layer the desired output of the nodes and thus the required change of the weights is obvious. \citet{rum} presented an algorithm based on the back-propagation of errors through the network which makes it possible to find $\frac{\partial E}{\partial w_{ij}}$ efficiently for all weights $w_{ij}$ in a multi-layer Perceptron. We are going to introduce this algorithm here, as it is also the foundation of the neural networks we use.

Using the chain rule we can rewrite the derivative
\begin{equation}
\frac{\partial E}{\partial w_{ij}} = \frac{\partial E}{\partial a_i}\frac{\partial a_i}{\partial w_{ij}} = \frac{\partial E}{\partial a_i} x_j 
\label{eqn:rewritederiv}
\end{equation}
in terms of derivatives with respect to nodes' inputs $a_i$ and their outputs or activations $x_i$.

Since the activations $x_j$ for any node can easily be found by feeding the respective data set to the network, the remaining task consists in calculating for each node $i$ the quantity $\delta_i$ often referred to as \emph{error}
\begin{equation}
\delta_i := \frac{\partial E}{\partial a_i} \; .
\label{eqn:deltadef}
\end{equation}

For the output layer of nodes with activation functions $f(a_i)=:f_i$, the derivative can be written as
\begin{equation}
\delta^{\mathrm{out}}_i = \frac{\partial E}{\partial a^{\mathrm{out}}_i} = \frac{\partial E}{\partial x^{\mathrm{out}}_i} \frac{\partial x^{\mathrm{out}}_i}{\partial a^{\mathrm{out}}_i} = \frac{\partial E}{\partial x^{\mathrm{out}}_i} f'^{\mathrm{out}}_i \; .
\label{eqn:outputdeltas}
\end{equation}

In the common case of identity activation functions for the output layer and squared error cost functions $E=\frac{1}{2}\sum_i (x^{\mathrm{out}}_i - \hat{x}_i)^2$ with true results $\hat{x}_i$ for single training sets this leads to the simple residuals of the output nodes' activations,
\begin{equation}
\delta^{\mathrm{out}}_i = x^{\mathrm{out}}_i - \hat{x}_i \; .
\label{eqn:deltao1}
\end{equation}

For a node $j$ from any other layer we can simplify the expression for $\delta_j$ using a sum over the nodes $k$ of the following layer, requiring that the activation functions $f_j$ be differentiable:
\begin{equation}
\delta_j = \frac{\partial E}{\partial a_j} = \sum_k \frac{\partial E}{\partial a_k} \frac{\partial a_k}{\partial a_j} = \sum_k \frac{\partial E}{\partial a_k} \frac{\partial a_k}{\partial x_j} \frac{\partial x_j}{\partial a_j} = f'_j\sum_k \delta_k w_{kj} \; .
\label{eqn:deltaprop}
\end{equation}

Thus with eqn. (\ref{eqn:deltaprop}) we can calculate one layer's $\delta$s using the derivative of its activations functions, the subsequent layer's $\delta$s and the weights of the connections between the two layers. The output layer's $\delta$s given in eqn. (\ref{eqn:outputdeltas}), we can back-propagate the errors through the network to find $\delta_i$ for each node. Finally, we can change the weights accordingly (using equations (\ref{eqn:graddesc}), (\ref{eqn:rewritederiv}) and (\ref{eqn:deltadef})):

\begin{equation}
\Delta w_{ij} = - \eta \frac{\partial E}{\partial w_{ij}} = - \eta \frac{\partial E}{\partial a_i} x_j = - \eta \delta_i x_j \; .
\label{eqn:bpweightcorrection}
\end{equation}

Applying these changes with a small enough parameter $\eta > 0$, it is possible to gradually minimize the cost function and therefore improve the network performance. Also, the gradient found can be used with a Quasi-Newton algorithm.

\subsection{Back-Propagation in Averaging Perceptrons}
\label{sec:bpap}
For averaging Perceptrons we define the cost function as the sum of squared errors of the output averages $\langle x^{\mathrm{out}} \rangle^i$ of a meta set $i$ of single data sets against known true average outputs $\hat{x}_i$ for the meta data set:
$$
E=\frac{1}{2}\sum_i (\langle x^{\mathrm{out}} \rangle^i - \hat{x}_i)^2 \; .
$$

We define errors $\delta$ for each node, this time referring to a single data set $m$ with unknown true result which is part of a meta data set with known average true result. Recalling definition (\ref{eqn:deltadef}) and the form of arithmetic averages we find
$$
\delta_j^m = \frac{\partial E}{\partial a_j^m} \; ,
$$
which simplifies for the output nodes to a constant error for all single data sets
$$
\delta_i^{o,m} = \frac{1}{n}(\langle x^{\mathrm{out}} \rangle^i - \hat{x}_i) \quad \forall m \; ,
$$
thus $\sum_m \delta_i^{\mathrm{out},m} = \langle x^{\mathrm{out}} \rangle^i - \hat{x}_i = \delta_i^{\mathrm{out}}$ as defined in eqn. (\ref{eqn:deltao1}) above.

For hidden nodes we find, in analogy to eqn. (\ref{eqn:deltaprop}) summing over nodes $k$ of the subsequent layer,
$$
\delta_j^m = f'^m_j \sum_k w_{kj} \delta_k^m \; ,
$$
which depends on $m$ due to the different activations and therefore activation derivatives of the nodes for each primitive data set. Note that $\sum_m \delta_j^m$ is no longer equal to $\delta_j$ as defined in eqn. (\ref{eqn:deltaprop}), as $f'^m_j$ and $\delta_k^m$ are not necessarily independent.

Calculating the derivative of the cost function with respect to each weight in turn we find the quantity needed for gradient descent:
$$
\Delta w_{ij} = -\eta \frac{\partial E}{\partial w_{ij}} = -\eta \sum_m \frac{\partial E}{\partial a_i^m}\frac{\partial a_i^m}{\partial w_{ij}}=-\eta\sum_m \delta_i^m x_j^m \; .
$$

An algorithm following this calculation therefore first has to find network averages for the meta data set under consideration. After having calculated the constant $\delta_i^{\mathrm{out}}$ from that, it has to feed each primitive data set $m$ in turn, back-propagate $\delta^m$ through the network and sum $\delta_i^m x_j^m$ for each weight $w_{ij}$.




\clearpage


\end{document}